\documentclass[%
aip,
sd,%
amsmath,amssymb,
onecolumn,
reprint,%
author-year,%
]{revtex4-1}

\usepackage{natbib}
\usepackage{graphicx}
\usepackage{dcolumn}
\usepackage{bm}
\usepackage{graphicx}
\usepackage{epsfig}

\usepackage{amssymb}
\usepackage{amsthm}
\usepackage{amsbsy}
\usepackage{amsmath}
\usepackage{amsfonts}
\usepackage{dsfont}
\usepackage{rotating}

\begin{document}

\preprint{AIP/123-QED}

\title[]
{High order numerical simulations of
the Richtmyer Meshkov instability in a relativistic fluid}

\author{O. Zanotti}
\email{Olindo.Zanotti@unitn.it}
\author{M. Dumbser}%
\email{Michael.Dumbser@unitn.it}
\affiliation{ 
$^1$
Laboratory of Applied Mathematics,
Via Mesiano 77, 38123, Trento, Italy
}%

\date{\today}

\begin{abstract}
We study the Richtmyer--Meshkov (RM) instability of a relativistic perfect fluid by means of high order numerical simulations with adaptive mesh refinement (AMR). 
The numerical scheme adopts a finite volume Weighted Essentially Non-Oscillatory (WENO) reconstruction to increase accuracy in space, 
a local space-time discontinuous Galerkin predictor method to obtain high order of accuracy in time 
and a high order one-step time update scheme together with a ``cell-by-cell'' space-time AMR strategy with time-accurate local time stepping. 
In this way, third order accurate (both in space and in time) numerical simulations of the RM instability are performed, spanning a wide parameter space. 
We present results both for the case in which a light fluid penetrates into a higher density one (Atwood number $A>0$), and for 
the case in which a heavy fluid penetrates into a lower density one (Atwood number $A<0$).
We find that, for large Lorentz factors $\gamma_{\rm s}$ of the incident shock wave, the relativistic RM instability is substantially weakened and ultimately suppressed.
More specifically, the growth rate of the RM instability in the linear phase has a local maximum which occurs at 
a critical value of $\gamma_{\rm s}\approx [1.2,2]$. 
Moreover, we have also revealed a genuine relativistic effect, absent in Newtonian hydrodynamics, which arises 
in three dimensional configurations with a non-zero velocity component tangent to the incident shock front.
In this case, the RM instability is strongly affected, typically resulting in less efficient mixing of the fluid. 
%
\end{abstract}

\maketitle


\newcommand{\be}{\begin{equation}}
\newcommand{\ee}{\end{equation}}
\newcommand{\bdm}{\begin{displaymath}}
\newcommand{\edm}{\end{displaymath}}
\newcommand{\bea}{\begin{eqnarray}}
\newcommand{\eea}{\end{eqnarray}}
\newcommand{\PNM}{P_NP_M}
\newcommand{\halb}{\frac{1}{2}}
\newcommand{\FQi}{\tens{\mathbf{F}}\left(\Qi\right)}
\newcommand{\FQj}{\tens{\mathbf{F}}\left(\Qj\right)}
\newcommand{\FQjj}{\tens{\mathbf{F}}\left(\Qjj\right)}
\newcommand{\nj}{\vec n_j}
\newcommand{\FORCE}{\textnormal{FORCE}}
\newcommand{\GFORCE}{\textnormal{GFORCEN}}
\newcommand{\LF}{\textnormal{LF}'}
\newcommand{\LW}{\textnormal{LW}'}
\newcommand{\WL}{\mathcal{W}_h^-}
\newcommand{\WR}{\mathcal{W}_h^+}
\newcommand{\nur}{\boldsymbol{\nu}^\textbf{r} }
\newcommand{\nuf}{\boldsymbol{\nu}^{\boldsymbol{\phi}} }
\newcommand{\nut}{\boldsymbol{\nu}^{\boldsymbol{\theta}} }
\newcommand{\ar}{\phi_1\rho_1}
\newcommand{\arr}{\phi_2\rho_2}
\newcommand{\ur}{u_1^r}
\newcommand{\uf}{u_1^{\phi}}
\newcommand{\ut}{u_1^{\theta}}
\newcommand{\urr}{u_2^r}
\newcommand{\uff}{u_2^{\phi}}
\newcommand{\utt}{u_2^{\theta}}
\newcommand{\ub}{\textbf{u}_\textbf{1}}
\newcommand{\ubb}{\textbf{u}_\textbf{2}}
\newcommand{\RoeMat}{{\tilde A}_{\Path}^G} 
\renewcommand{\u}{\mathbf{u}}
\newcommand{\q}{\mathbf{q}}
\newcommand{\w}{\mathbf{w}}

\section{Introduction}
\label{sec:level1}
The Richtmyer--Meshkov (RM) instability is a typical fluid instability, predicted analytically by \cite{Richtmyer1960} and 
discovered experimentally by \cite{Meshkov1969}, which develops when a shock wave 
crosses a contact discontinuity within a fluid, or between two different fluids [see \cite{Brouillette2002} for a review]. 
Under terrestrial physical conditions, the RM instability is, for instance, encountered in 
inertial confinement fusion, where shock waves produced by beams of laser hit a small capsule containing a deuterium-tritium fuel. Typically, the instability causes the premature mixing of the fuel with the outer shell and
represents a serious obstacle in the process of energy extraction~\citep{Taylor1997,Amendt2002,Wilson2004b}. 
The RM instability is also encountered, though under rather different physical conditions, in a few astrophysical systems.
A first example is given by supernova remnant formation,  in which  the 
reflected shock that travels through the interaction region may activate the RM instability and account for the 
mixing of the inner and outer layers of the progenitor star \citep{Chevalier1992,Kane1999,Abarzhi2005,Kifonidis2006}. 
A second example is given by relativistic jets, whose transverse structure can be substantially affected by the development of the RM instability acting at the interface between the jet and the surrounding medium~\citep{Matsumoto2013}.

While in inertial confinement fusion the shock speeds involved are well below the speed of light $c$, being of the order of $\sim 10^{-4} c$~\citep{Holmes1999}, in astrophysical systems, and in particular for supernovae,
the speed of the incident shock can approach $c$, making a relativistic treatment necessary. 
Unlike the Newtonian version of the RM instability, which has been studied in great detail both in the linear and in the nonlinear regime by means of theoretical analysis and numerical simulations [see, among the others, \cite{Kull1991,Dimonte1996,Velikovich1996,Mikaelian1996,Mikaelian1998,Kotelnikov2000,Wouchuk2001,Thornber2012}], the special relativistic version of the RM instability has been considered only in a few recent works and relatively little is known about it.
\cite{Mohseni2013}, for instance, after performing two and three dimensional numerical simulations, found that, above a critical value of the fluid velocity, relativistic effects produce a decrease in the growth rate of the instability. \cite{Inoue2012}, on the other hand, studied the RM instability of a current sheet in a relativistic plasma with magnetic field.
In this paper we extend the numerical analysis of \cite{Mohseni2013} by exploring a wide parameter space.
In particular, we consider the dependence of the growth rate of the instability on the Lorentz factor $\gamma_{\rm s}$ of the shock wave, up to
$\gamma_{\rm s}=10$ and we investigate the impact 
on the RM instability of a non-zero velocity component tangential to the shock front.  

In order to obtain  reliable quantitative predictions about the relativistic RM instability, the choice of the numerical method
turns out to be absolutely crucial.
Due to the nature of the problem, which involves the propagation of strong shocks, the so-called   
{\em Godunov methods}, which are based on the conservative formulation of the equations, should be preferred with respect
to any other numerical method.\footnote{Numerical methods not based on the conservative formulation of the equations do, in general, \textit{not} provide the correct 
propagation speed of discontinuities \citep{Leveque92,HouLeFloch}, with very harmful consequences on the study of the RM instability.}
Their application  to the solution of the special relativistic hydrodynamics equations
dates back to \citet{Marti91}, who introduced them with special attention to astrophysical systems. 
Moreover, the numerical modeling of complex flow structures and fluid instabilities would also benefit significantly if {\em high order numerical schemes} are adopted. 
The first implementation of a Godunov method, for relativistic hydrodynamics, 
with an order of convergence higher than the second was obtained  by \citet{DelZanna2002}, who used a finite difference 
approach with Essentially Non-Oscillatory (ENO) reconstruction in space and a  Runge--Kutta integration in time. 
In this paper we follow an entirely different strategy, proposed by  \cite{DET2008}, which is based on the combination of a few
specific steps, some of which are classical, while some others are highly innovative. 
On one hand, we adopt a standard finite volume scheme, evolving in time the cell-averages of the conservative quantities. On the other hand, by relying on the weak integral 
form of the governing partial differential equations (PDEs), we compute, locally for each cell and before the solution of the Riemann problem at each cell interface, the 
evolution in time of the polynomial obtained after the reconstruction through the Weighted Essentially Non-Oscillatory (WENO) method. In this way an arbitrary high order 
numerical scheme with just one step for the time update can be obtained. The last crucial component of our numerical strategy is represented by the implementation of 
adaptive mesh refinement (AMR), which is particularly convenient when complex flows structures form in limited portions of the numerical grid. 
The general approach has been presented very recently by \citet{Dumbser2012b} and \citet{Zanotti2013}, while here we apply it for the first time to a relevant physical 
problem of relativistic fluid dynamics. 

In the rest of the paper we set the speed of light $c=1$.

\section{Mathematical framework}
\label{sec:level2}

\subsection{The relativistic hydrodynamics equations}
In the following we neglect gravity, and we therefore assume a 
flat space-time in Cartesian coordinates, whose metric is simply given by
\be
\mathrm{d}s^2=g_{\mu\nu}dx^\mu dx^\nu=-\mathrm{d}t^2+\mathrm{d}x^2+\mathrm{d}y^2+\mathrm{d}z^2\,.
\ee
Moreover, we limit our attention to perfect fluids, which are described by an energy momentum tensor $T^{\alpha\beta}$
\be
T^{\alpha\beta}=\rho h\,u^{\,\alpha}u^{\beta}+pg^{\,\alpha\beta}\,,
\label{eq:T_matter}
\ee
where $u^\alpha$ is
the four-velocity of the fluid, while $\rho$, $h=1+\epsilon + p/\rho$,
$\epsilon$ and $p$ are the rest-mass density, the specific (per unit mass) enthalpy,
the specific internal energy, and the thermal pressure,
respectively. The equation of state is that of an ideal-gas, i.e.
\be
p=\rho\epsilon(\Gamma-1) \,,
\ee
where $\Gamma=4/3$ is the adiabatic index of the gas. 
The resulting fluid is therefore relativistic both in the equation of state and in the dynamics.
The mathematical structure of the equations of inviscid special relativistic hydrodynamics
is similar to that of its Newtonian analog. In fact, just like the Euler equations of gas dynamics, they can be written 
as a hyperbolic system of conservation laws~\citep{Marti91}, i.e. 
\be
\partial_t{\bf u} + \partial_i{\bf f}^i=0\,,
\label{eq:UFS}
\ee
where the conservative variables and the correspondent
fluxes in the $i$ direction are given by
\be
{\bf u}=\left[\begin{array}{c}
D \\ S_j \\ E
\end{array}\right],~~~
{\bf f}^i=\left[\begin{array}{c}
 v^i D \\
 W^i_j \\
 S^i
\end{array}\right]\,.
\label{eq:fluxes}
\ee
The {\em conserved variables} $(D,S_j,E)$ can be written in terms
of  the rest-mass density $\rho$, of the thermal
pressure $p$ and of the fluid velocity $v_i$ by
\bea
&&D   = \rho \gamma ,\\
&&S_i = \rho h \gamma^2 v_i, \\
&&E   = \rho h \gamma^2 - p, 
\label{eq:cons}
\eea
where $\gamma=(1-v^2)^{-1/2}$ is the Lorentz factor of the fluid
with respect to the laboratory observer and
\be
W_{ij} \equiv \rho h \gamma^2 v_i v_j +p \delta_{ij} \\
\label{eq:W} 
\ee
is the fully spatial projection of the energy-momentum
tensor of the fluid. We emphasize that the recovering of the so-called {\em primitive variables}
$(\rho,v_i,p)$ from the conserved variables cannot be obtained in a closed form, and it typically requires the
solution of an algebraic equation.
The mathematical properties of the system of 
equations (\ref{eq:UFS}), including the computation
of the corresponding Jacobian
matrix, its eigenvectors and
eigenvalues, have been investigated
deeply over the years, and can be found in
\citet{Rezzolla_book:2013}.

%

\subsection{The numerical method}
\label{sec:NumMethod}
The special relativistic hydrodynamics equations expressed by Eq.~(\ref{eq:UFS}) 
form a hyperbolic system of conservation laws. As such, they can be solved by resorting to the
wide family of Godunov methods, based on Riemann solvers~\citep{Toro99}. 
A fundamental step-forward in the development of high order Godunov numerical schemes for conservation laws was obtained by
\cite{DET2008}, while the inclusion of AMR has been considered by \citet{Dumbser2012b} and \citet{Zanotti2013}.
The overall idea, whose details can be found in the above mentioned works,
can be summarized as follows. First of all, a traditional finite volume approach is adopted, according to which
the cell averages of the conservative variables, namely
\begin{equation}
{\bf \bar u}_{ijk}^{n}=\frac{1}{\Delta x_i}\frac{1}{\Delta y_j}\frac{1}{\Delta z_k}\int \limits_{x_{i-\halb}}^{x_{i+\halb}}\int \limits_{y_{j-\halb}}^{y_{j+\halb}}\int \limits_{z_{k-\halb}}^{z_{k+\halb}}{ \bf u}(x,y,z,t^n)dz\,\,dy\,\,dx
\end{equation}
define the conserved variables to be evolved in time, where 
$I_{ijk}=[x_{i-\halb};x_{i+\halb}]\times[y_{j-\halb};y_{j+\halb}]\times[z_{k-\halb};z_{k+\halb}]$ is the control volume at time $t^n$.
Second, given a stencil containing a prescribed number of elements, a nonlinear WENO reconstruction is performed in order to obtain a high order 
polynomial $\w_h(x,y,z,t^n)$ from the known cell averages, which approximates the solution within each cell at time $t^n$. Third,  
a local space--time Discontinuous Galerkin predictor 
based on the \textit{weak} integral form of the governing PDE is used to obtain the time evolution of the reconstructed polynomials 
inside each control volume. The resulting space-time polynomial is denoted by $\q_h(x,y,z,t)$. Having done that, a high order accurate 
computation of the numerical fluxes between adjacent elements is obtained by resorting to traditional Riemann solvers as
\begin{equation}
{\bf \tilde f}_{{\rm RP}}={\bf \tilde f} \! \left({\bf q}_h^-(x_{i+\halb},y,z,t),{\bf q}_h^+(x_{i+\halb},y,z,t)\right)\,,
\end{equation}
where ${\bf q}_h^-$ and ${\bf q}_h^+$ are the left and 
right boundary extrapolated states at the interface between adjacent elements, while  the functional form of
${\bf \tilde f}_{{\rm RP}}$ depends on the specific choice of the Riemann solver adopted~\citep{Toro99}. 
In all the simulations reported in this work, we have used the popular
HLL Riemann solver by \cite{Harten83}, which does not rely on the
characteristic structure of the equations and only needs the knowledge of the fastest and of the slowest eigenvalue.
Once the numerical fluxes have been computed, 
a one-step time-update scheme, with no need for intermediate Runge--Kutta stages, can be implemented as
\begin{eqnarray}
\label{eq:finite_vol}
{\bf \bar u}_{ijk}^{n+1}={\bf \bar u}_{ijk}^{n}&-&\frac{\Delta t}{\Delta x_i}\left({\bf f}_{i+\halb,jk}-{\bf f}_{i-\halb,jk} \right)-\frac{\Delta t}{\Delta y_j}\left({\bf g}_{i,j+\halb,k}-{\bf g}_{i,j-\halb,k} \right)\\
\nonumber
& -&\frac{\Delta t}{\Delta z_k}\left({\bf h}_{ij,k+\halb}-{\bf h}_{ij,k-\halb} \right)
\end{eqnarray}
where
\begin{equation}
\label{flux:F}
{\bf f}_{i+\halb,jk}= \frac{1}{\Delta t}\frac{1}{\Delta y_j}\frac{1}{\Delta z_k} \hspace{-1mm}  \int \limits_{t^n}^{t^{n+1}} \! \int \limits_{y_{j-\halb}}^{y_{j+\halb}}\int \limits_{z_{k-\halb}}^{z_{k+\halb}} \hspace{-1mm} 
{\bf \tilde f}_{{\rm RP}}(x_{i+\halb},y,z,t) \, dz \, dy \, dt\,,
\end{equation}
(and similarly for ${\bf g}_{i,j+\halb,k}$ and ${\bf h}_{ij,k+\halb}$) are the space--time averaged numerical fluxes.
All these ideas can be combined with the AMR approach, which we have implemented according to a "cell-by-cell'' refinement criterion 
\citep{Khokhlov1997} and parallelized through the standard Message Passing Interface (MPI) paradigm (see \citet{Zanotti2013} for all the details).

\section{Analysis of the RM instability}
\label{sec:level3}
%
\subsection{Initial conditions}
\label{sec:IC}

\begin{figure}
\begin{center}
{\includegraphics[angle=0,width=12.0cm,height=2.0cm]{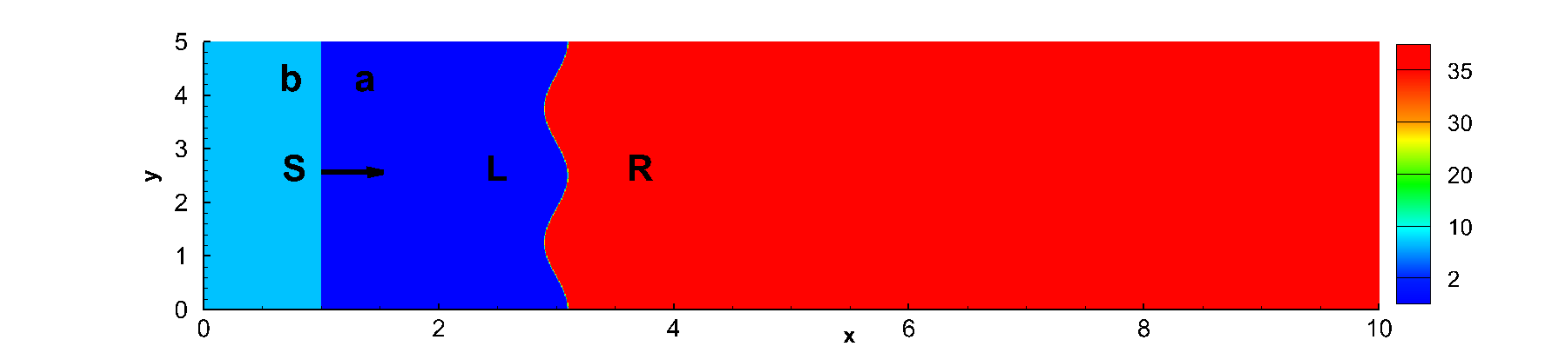}}
\end{center}
\caption{Schematic representation of the initial conditions in a representative model with Atwood number $A>0$. 
A right propagating shock wave, indicated by the black arrow, 
approaches the sinusoidal perturbation in the density located at $x=3$. The labels "a'' and "b'' stand for "ahead" and "behind'' of the shock front, 
respectively.
The labels "L'' and "R'' indicate the regions to the left and to the right of the initial perturbation. 
}
\label{fig:IC}
\end{figure}

We have studied the RM instability, neglecting gravity effects, by performing special relativistic hydrodynamics simulations in two and in three spatial dimensions 
with the method described in Sect.~\ref{sec:NumMethod}. 
Fig.~\ref{fig:IC} shows a schematic representation of the initial conditions for one representative model in the two-dimensional case. In our calculations, we always assume that the $x$ axis is oriented along the direction of propagation of the shock wave.
The perfect fluid is initially at rest ($v^x=0,v^y=0$), with a jump in the density between the two states "L'' and "R'', which defines a corresponding
pre-shock Atwood number $A=(\rho_R-\rho_L)/(\rho_R+\rho_L)$.
The two states are separated by a sinusoidal perturbation at $x_0+a\sin(\pi/2+2\pi y/\lambda)$, where $x_0=3$, $a=0.1$ and $\lambda=2.5$. 
At time $t=0$ a single shock wave, marked with "S'' in the figure, is placed at $x=1$ and propagates towards the right with prescribed Lorentz factor $\gamma_{\rm s}$.
The shock wave is built after solving the relativistic Rankine--Hugoniot conditions~\citep{Taub1948}, which, for our specific set-up, can be rephrased as follows. We first adopt the standard convention by which the
difference of a quantity evaluated ahead (subscript $a$) and behind
(subscript $b$) of the shock wave is denoted as $[[F]] \equiv F_a - F_b$.
The hydrodynamical quantities ahead of the shock are all assigned by the state "L''.
On the contrary, we keep the freedom to change the pressure behind the shock, in such a way that the shock speed is tuned to the desired value.
\begin{table}
\caption{\label{tab:table1}
Description of the models. The columns report the name of the model, the rest-mass density, velocity and pressure in the states "L'' and "R'' at either sides of the perturbation, the pressure $p_b$ behind the right-propagating shock wave,
the velocity $V_{\rm s}$ of the shock wave, the corresponding Lorentz factor $\gamma_{\rm s}=1/\sqrt{1-V_{\rm s}^2}$ and the corresponding Mach number ${\cal M}_{\rm s}$. 
The adiabatic index is $\Gamma=4/3$ for all the models.
}
\vspace{0.3cm}
\begin{ruledtabular}
\begin{tabular}{l|lcl|lcl|llll}
Model  & $\rho_L$ & $\vec{v}_L$ & $p_L$ & $\rho_R$ & $\vec{v}_R$ & $p_R$ & $p_b$ & $V_{\rm s}$ & $\gamma_{\rm s}$ & ${\cal M}_{\rm s}$ \\
\hline
2D-Ia   & 1.0 & 0 & 0.1 & 35.0 & 0 & 0.1 & 0.23 & 0.416 & 1.1 & 1.45  \\
2D-Ib   & 1.0 & 0 & 0.1 & 35.0 & 0 & 0.1 & 1.35 & 0.745 & 1.5 & 3.44  \\
2D-Ic   & 1.0 & 0 & 0.1 & 35.0 & 0 & 0.1 & 2.40 & 0.831 & 1.8 & 4.61  \\
2D-Id   & 1.0 & 0 & 0.1 & 35.0 & 0 & 0.1 & 3.20 & 0.866 & 2.0 & 5.34  \\
2D-Ie   & 1.0 & 0 & 0.1 & 35.0 & 0 & 0.1 & 8.34 & 0.943 & 3.0 & 8.74  \\
2D-If   & 1.0 & 0 & 0.1 & 35.0 & 0 & 0.1 & 24.2 & 0.979 & 5.0 & 15.12 \\
\hline
2D-IIa   & 35.0 & 0 & 0.1 & 1.0 & 0 & 0.1 & 0.64  & 0.140 & 1.01 & 2.39  \\
2D-IIb   & 35.0 & 0 & 0.1 & 1.0 & 0 & 0.1 & 3.08  & 0.305 & 1.05 & 5.21   \\
2D-IIc   & 35.0 & 0 & 0.1 & 1.0 & 0 & 0.1 & 6.30  & 0.416 & 1.1  & 7.46   \\
2D-IId   & 35.0 & 0 & 0.1 & 1.0 & 0 & 0.1 & 20.39 & 0.639 & 1.3  & 13.52  \\
2D-IIe   & 35.0 & 0 & 0.1 & 1.0 & 0 & 0.1 & 36.43 & 0.745 & 1.5  & 18.19  \\
2D-IIf   & 35.0 & 0 & 0.1 & 1.0 & 0 & 0.1 & 85.03 & 0.866 & 2.0  & 28.18  \\
\hline
3D-Ia   & 1.0 &         0 & 0.1 & 35.0 &         0 & 0.1 & 1.35  & 0.74 & 1.5 & 3.44  \\
3D-Ib   & 1.0 & (0,0.9,0) & 0.1 & 35.0 & (0,0.9,0) & 0.1 & 6.87  & 0.74 & 1.5 & 3.44  \\
\hline
3D-IIa   & 35.0 &         0 & 0.1 & 1.0 &         0 & 0.1 & 6.30  & 0.42 & 1.1 & 7.46  \\
3D-IIb   & 35.0 & (0,0.9,0) & 0.1 & 1.0 & (0,0.9,0) & 0.1 & 32.30  & 0.42 & 1.1 & 7.46  \\
\end{tabular}
\end{ruledtabular}
\end{table}

In fact, from $p_b$ we compute $h_b$ through the Taub-adiabat as \citep{Pons00}
\begin{equation}
\label{Taub2}
\left[1+\frac{(\Gamma-1)(p_a-p_b)}{\Gamma p_b}\right]h_b^2
	-\frac{(\Gamma-1)(p_a-p_b)}{\Gamma p_b}h_b 
	+\frac{h_a (p_a-p_b)}{\rho_a}-h_a^2 = 0 \,.
\end{equation}
Having done that, the mass flux $J$ through the shock, which is invariant under Lorentz transformations along the $x-$ direction, is given by
\begin{equation}
\label{flux2}
J^2=-\frac{\Gamma}{\Gamma-1}\frac{[[p]]}{[[h(h-1)/p]]} \,,
\end{equation}
while the shock velocity and the fluid velocity behind the shock are 
\begin{eqnarray}
V_{\rm s} &=& \frac{ \rho_a^2 \gamma_a^2 v^x_a  +   
	|J| \sqrt{J^2 + \rho_a^2 \gamma_a^2 [1 -(v_a^x)^2]}}
            { \rho_a^2 \gamma_a^2 + J^2 }\,,  \\
						v_b^x &=& \frac{h_a \gamma_a v_a^x + \gamma_{\rm s}(p_b - p_a)/J}{h_a \gamma_a +
	(p_b - p_a)[\gamma_{\rm s} v_a^x/J + 1/(\rho_a \gamma_a)]}
	\,.
\label{velshock}
\end{eqnarray}
%
The relations above provide all the necessary information to specify the initial conditions 
of the RM instability in two-spatial dimensions.
We have considered either the situation when $A>0$ ({\em light-to-heavy} models)
and when  $A<0$ ({\em heavy-to-light} models).

In the three-dimensional case, on the other hand, it may also be interesting to investigate the effects of 
a velocity component along the $y$ (or the $z$) direction.
Unlike Newtonian hydrodynamics, the two velocities $v^y$ and $v^z$ are not continuous across the shock front
and the Rankine--Hugoniot conditions provide~\citep{Pons00}
\begin{equation}
\label{eq:Pons}
v_b^{y,z}=h_a \gamma_a v_a^{y,z}\sqrt{\frac{1-(v_b^x)^2}{h_b^2+(h_a \gamma_a v_a^{y,z})^2}}\,.
\end{equation}
The tangential velocities, which enter through the Lorentz factor $\gamma_a$, have already been shown to produce 
genuinely new physical effects in relativistic hydrodynamics, such as in the solution of the Riemann problem, where they can be responsible of wave-pattern changes~\citep{Rezzolla02}. 
For these reasons, it is meaningful to investigate possible 
effects of tangential velocities also in the relativistic three-dimensional RM instability.

Table~\ref{tab:table1} contains the most relevant physical parameters of the models that we have studied. We recall that the relativistic Mach number of the shock wave, 
denoted by ${\cal M}_{\rm s}$ in the table, can be computed as ${\cal M}_{\rm s}= (V_{\rm s}/c_{a}) (\gamma_{\rm s}/\gamma_{c_a})$, where 
$c_{a}$ is the sound speed in the un-shocked fluid, while $\gamma_{c_a}=1/\sqrt{1-c_{a}^2}$ is the corresponding Lorentz factor. 
In the following we present our results obtained
with a finite volume scheme at the third order of accuracy, both in space and in time. We emphasize that
the WENO reconstruction has been performed on the characteristic variables~\citep{Toro99}.

\subsection{Two dimensional simulations}
%
\begin{figure}
\begin{center}
{\includegraphics[angle=0,width=12.0cm,height=1.9cm]{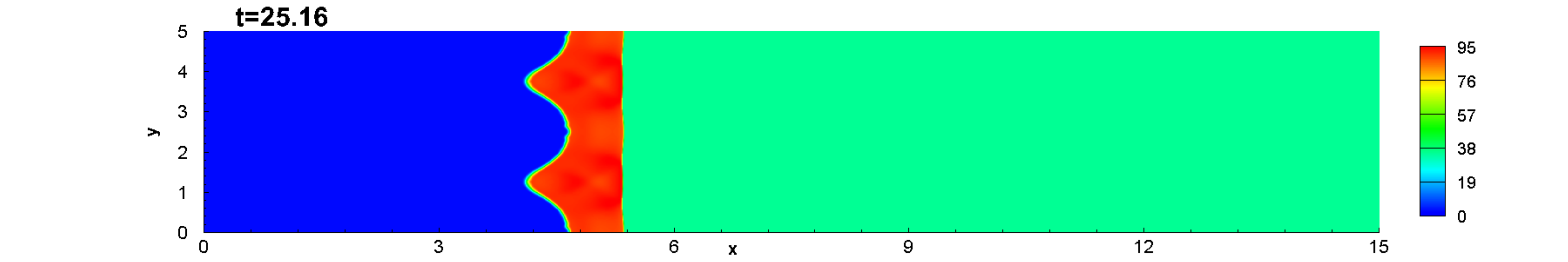}}
{\includegraphics[angle=0,width=12.0cm,height=1.9cm]{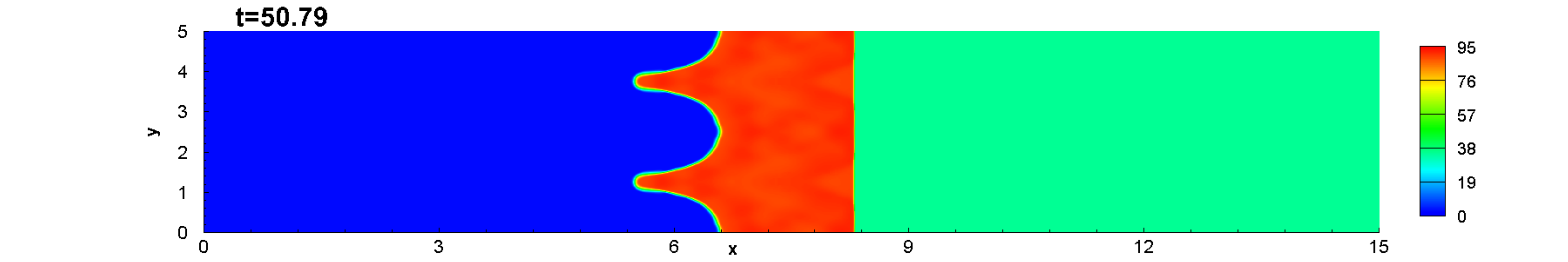}}
{\includegraphics[angle=0,width=12.0cm,height=1.9cm]{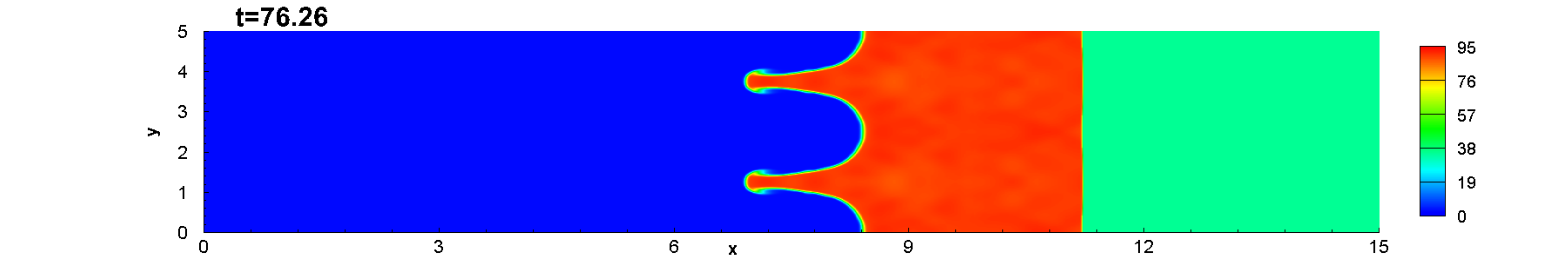}}
{\includegraphics[angle=0,width=12.0cm,height=1.9cm]{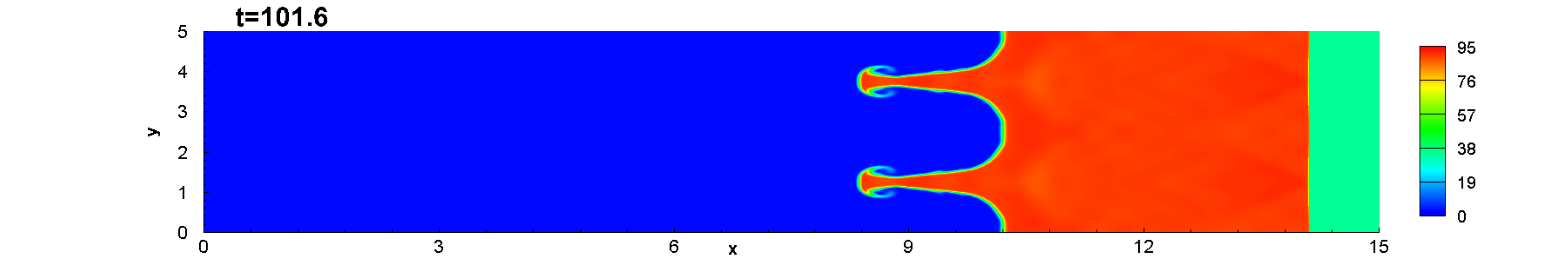}}
{\includegraphics[angle=0,width=12.0cm,height=1.9cm]{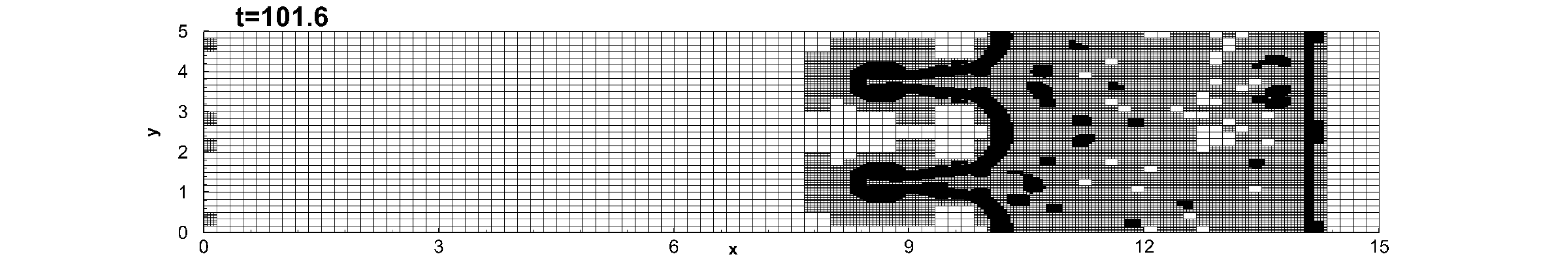}}
\end{center}
\caption{{\em Light-to-Heavy}. Time evolution of the RM instability in a configuration with Lorentz factor $\gamma_{\rm s}=1.1$ and $A=0.94$.
At time $t=101.6$ the AMR grid is composed of $114,576$ elements.
}
\label{fig:RMW1.1}
\end{figure}
%
\begin{figure}
\begin{center}
{\includegraphics[angle=0,width=12.0cm,height=1.9cm]{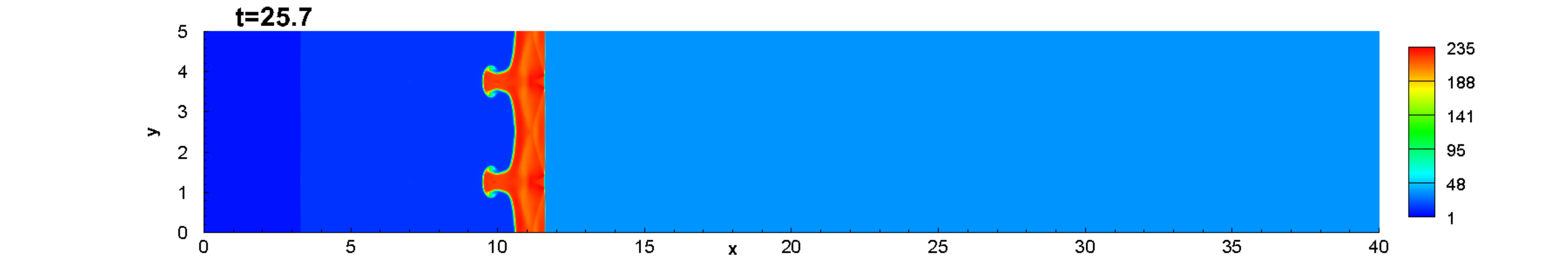}}
{\includegraphics[angle=0,width=12.0cm,height=1.9cm]{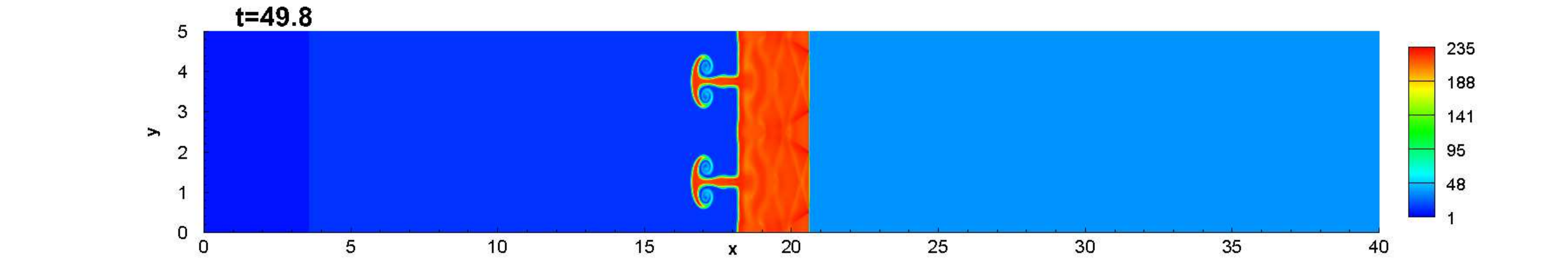}}
{\includegraphics[angle=0,width=12.0cm,height=1.9cm]{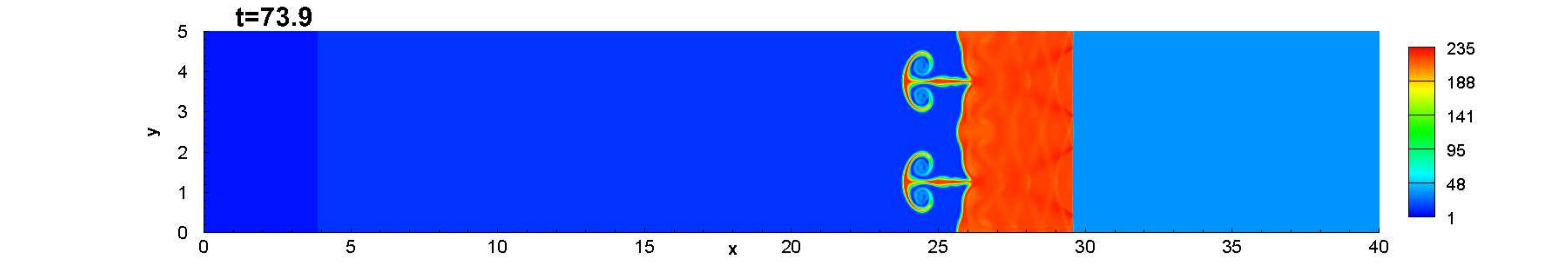}}
{\includegraphics[angle=0,width=12.0cm,height=1.9cm]{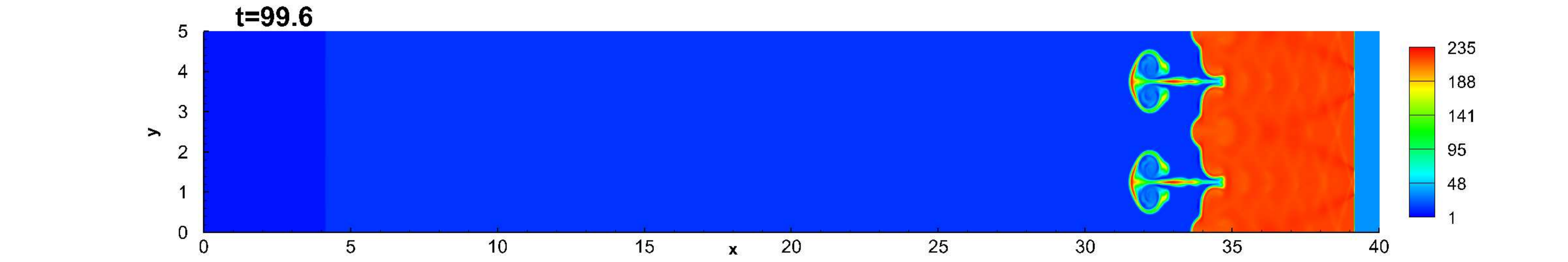}}
{\includegraphics[angle=0,width=12.0cm,height=1.9cm]{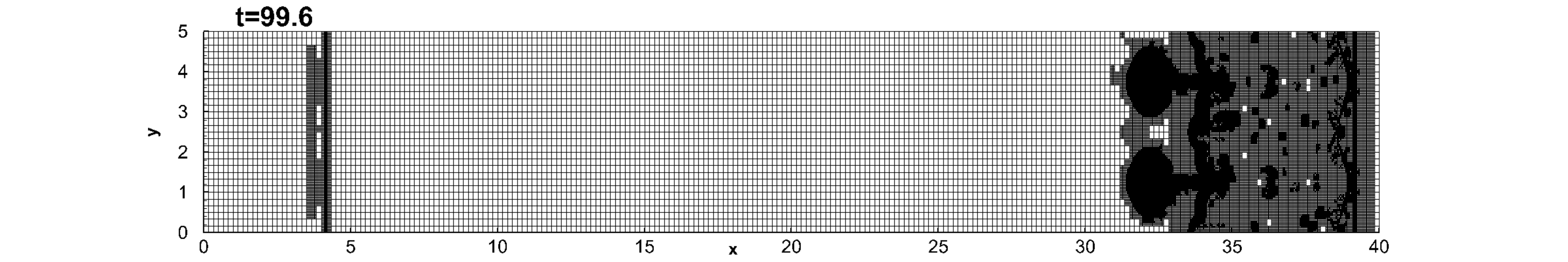}}
\end{center}
\caption{{\em Light-to-Heavy}. Time evolution of the RM instability in a configuration with Lorentz factor $\gamma_{\rm s}=1.5$ and $A=0.94$.
At time $t=99.6$ the AMR grid is composed of $266,944$ elements.
}
\label{fig:RMW1.5}
\end{figure}
\begin{figure}
\begin{center}
{\includegraphics[angle=0,width=12.0cm,height=1.9cm]{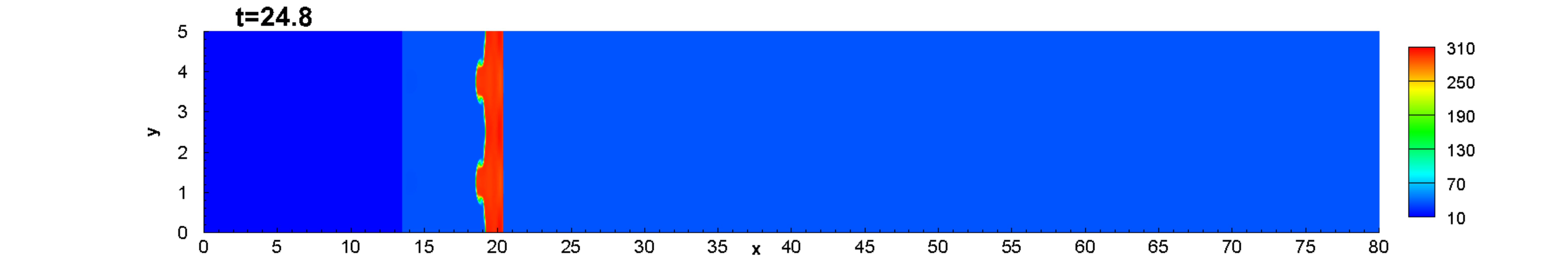}}
{\includegraphics[angle=0,width=12.0cm,height=1.9cm]{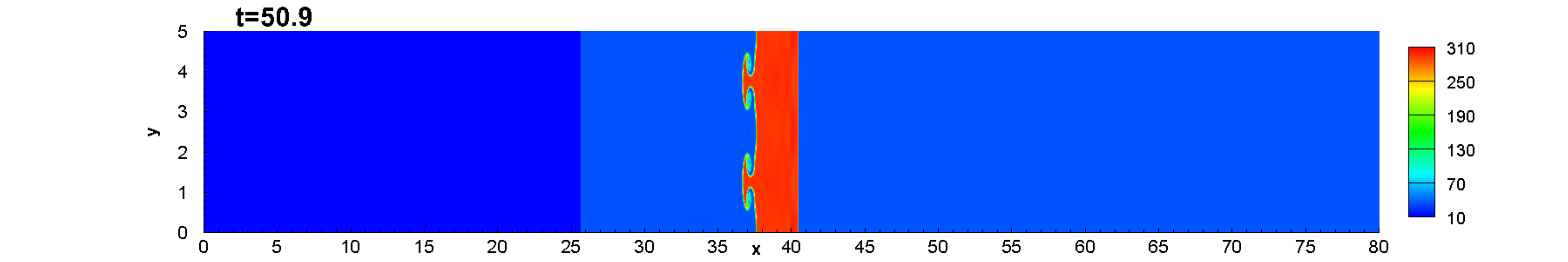}}
{\includegraphics[angle=0,width=12.0cm,height=1.9cm]{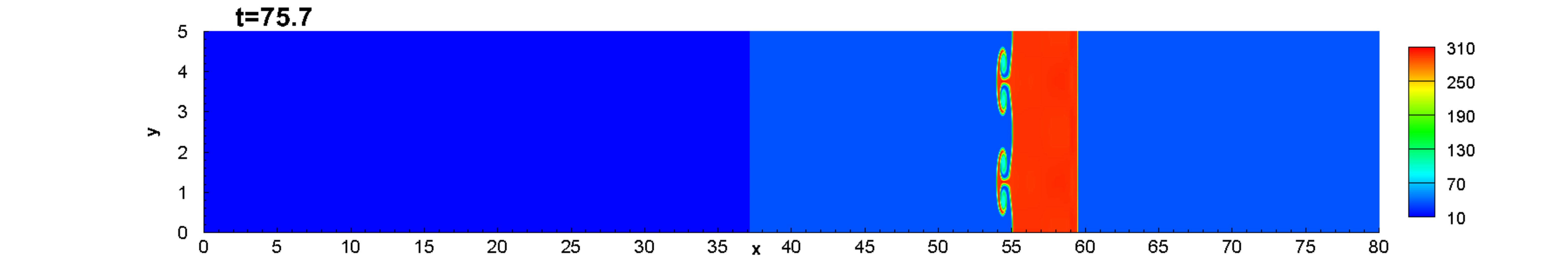}}
{\includegraphics[angle=0,width=12.0cm,height=1.9cm]{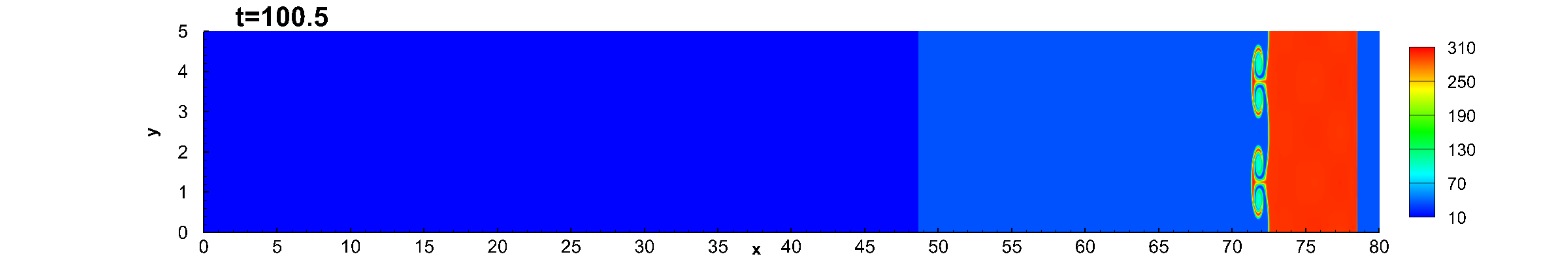}}
{\includegraphics[angle=0,width=12.0cm,height=1.9cm]{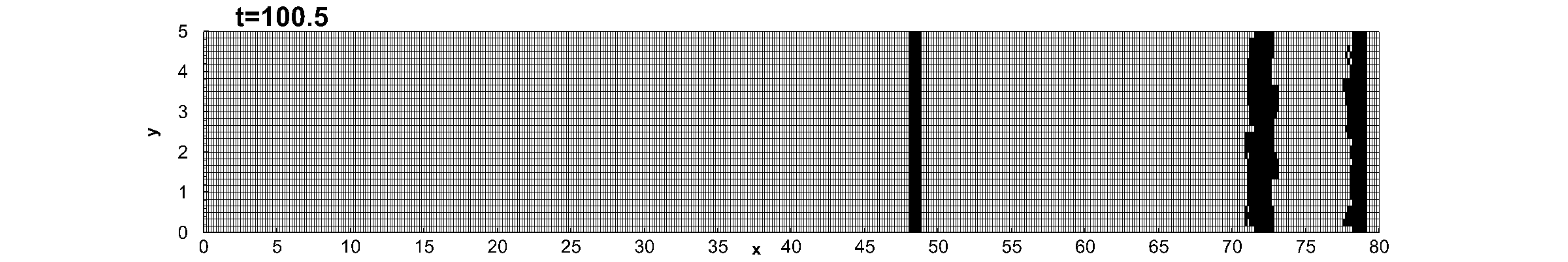}}
\end{center}
\caption{{\em Light-to-Heavy}. Time evolution of the RM instability in a configuration with Lorentz factor $\gamma_{\rm s}=3$ and $A=0.94$.
At time $t=100.5$ the AMR grid is composed of $121,856$ elements.
}
\label{fig:RMW3}
\end{figure}
The computational domain is $\Omega=[0,100]\times[0,5]$, and it is initially covered by 
a level zero grid with $600 \times 30$ control volumes. The refinement factor of the AMR algorithm is $\mathfrak{r}=4$, 
and two levels of refinement have been activated. This corresponds to an equivalent resolution on a uniform fine 
grid of $4,608,000$ cells. Periodic boundary conditions are applied in the $y$ direction. 

We have studied the development of the RM instability in two different classes of models. The first class corresponds to the case
when a light fluid penetrates into a heavy fluid, i.e. {\em light-to-heavy} models with Atwood number $A>0$ (models
denoted as 2D-I in Table~\ref{tab:table1}),
while the second class considers the opposite situation, i.e. {\em heavy-to-light} models with $A<0$
(models denoted as 2D-II in Table~\ref{tab:table1}).

\subsubsection{Light-to-heavy models, $A>0$}
\label{sec:Light-to-heavy models}

\begin{figure}
\begin{center}
{\includegraphics[angle=0,width=7.0cm,height=7.0cm]{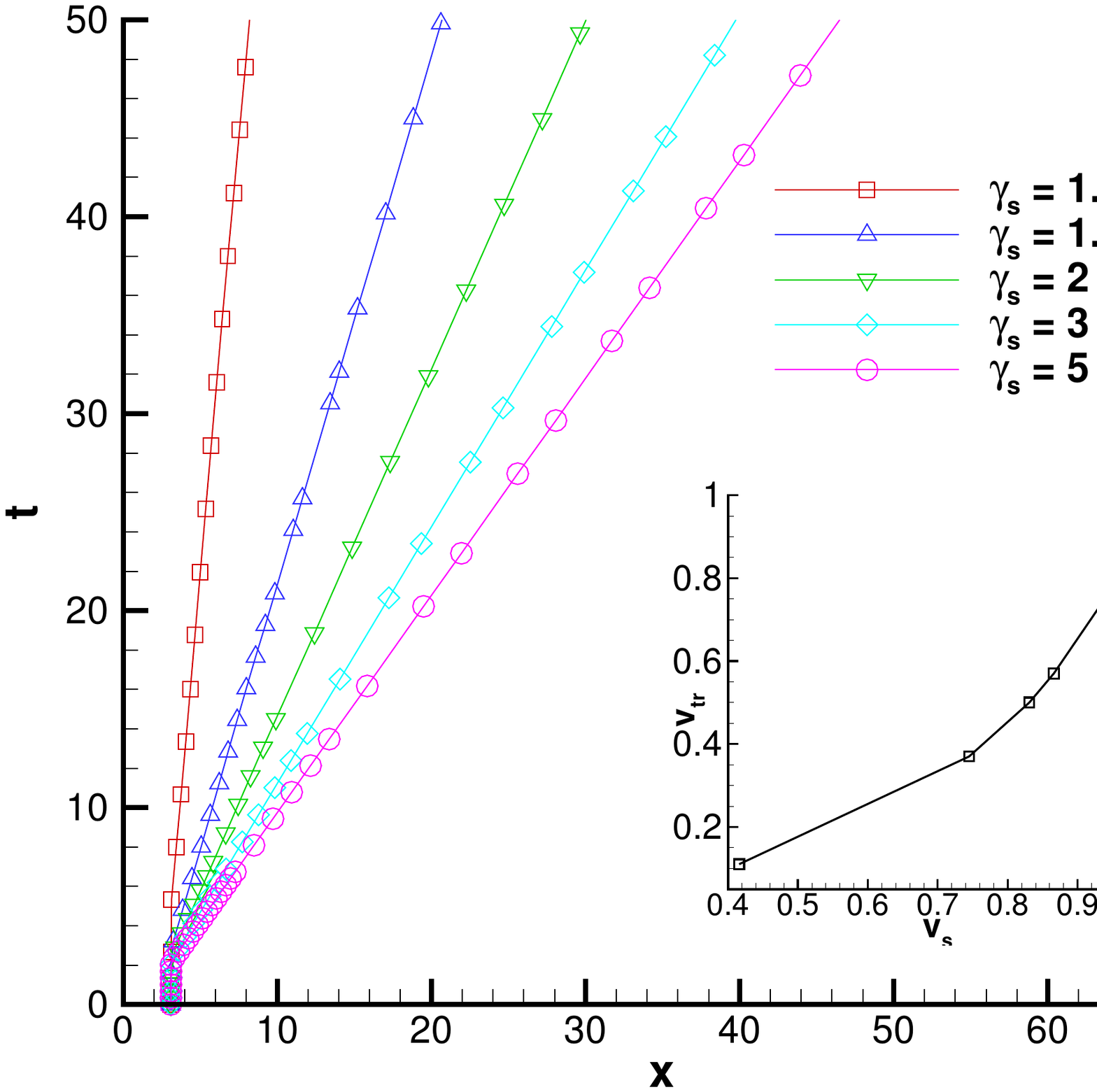}}
\end{center}
\caption{
{\em Light-to-Heavy}.
World lines of the transmitted shock waves for different  Lorentz factors $\gamma_{\rm s}$ of the incident shock wave.
The inset shows the non-linear grow of $v_{\rm tr}$ as the speed  $V_{\rm s}$ of the incident shock wave increases.
}
\label{fig:SpeedOfTransmittedShock}
\end{figure}
Figures \ref{fig:RMW1.1}--\ref{fig:RMW3} illustrate the dynamics of the RM instability for different velocities of the incident shock wave, namely for  $\gamma_{\rm s}=1.1$, $\gamma_{\rm s}=1.5$ and $\gamma_{\rm s}=3$, respectively.
Each figure is composed of five panels: the four top panels are snapshots at four times in the range $ t \in [0,\sim 100]$, while the bottom panel shows the AMR numerical grid corresponding to the last snapshot. After the initial shock wave hits the sinusoidal interface, a strong shock wave is transmitted into the high density region with a speed $v_{\rm tr}$ that is related to that of the incident shock (see below).
On the back of the shocked region, the sinusoidal interface is also dragged towards the right, but at a velocity slower than that of the transmitted shock. As a result, the size of the shocked region increases with time. 
However, this effect is reduced as the Lorentz factor $\gamma_{\rm s}$ increases, and indeed the size of the shocked region is much smaller 
for the model shown in 
Fig.~\ref{fig:RMW3} (with $\gamma_{\rm s}=3$) than it is for that in Fig.~\ref{fig:RMW1.1} (with $\gamma_{\rm s}=1.1$). 
In Fig.~\ref{fig:SpeedOfTransmittedShock} we have reported the world-lines of the transmitted shock-fronts in the spacetime diagram $(x,t)$.
The speeds of the transmitted shocks are constant and they are $v_{\rm tr}\sim 0.11$, $v_{\rm tr}\sim 0.37$, $v_{\rm tr}\sim 0.57$, $v_{\rm tr}\sim 0.76$ and $v_{\rm tr}\sim 0.91$ for
$\gamma_{\rm s}=1.1$, $\gamma_{\rm s}=1.5$, $\gamma_{\rm s}=2$, $\gamma_{\rm s}=3$ and $\gamma_{\rm s}=5$, respectively. The inset of Fig.~\ref{fig:SpeedOfTransmittedShock} highlights how
the speeds of the transmitted shocks grow non-linearly with $V_{\rm s}$.

\begin{figure}
\begin{center}
{\includegraphics[angle=0,width=12.0cm,height=1.9cm]{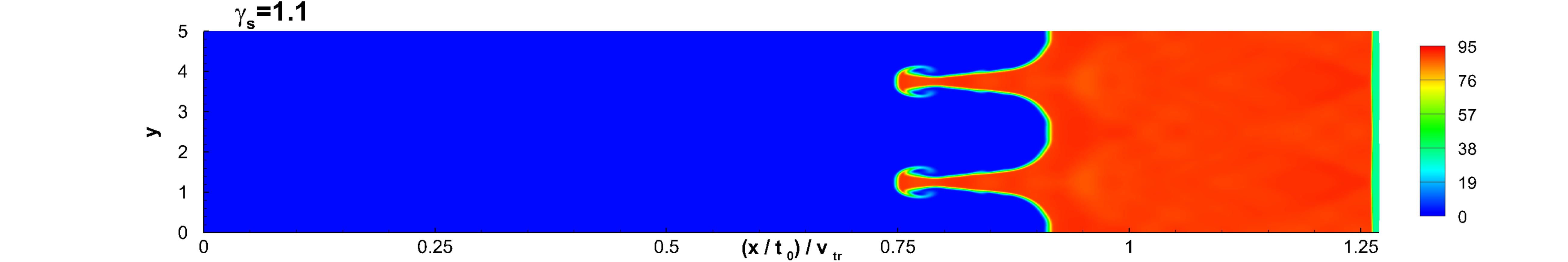}}
{\includegraphics[angle=0,width=12.0cm,height=1.9cm]{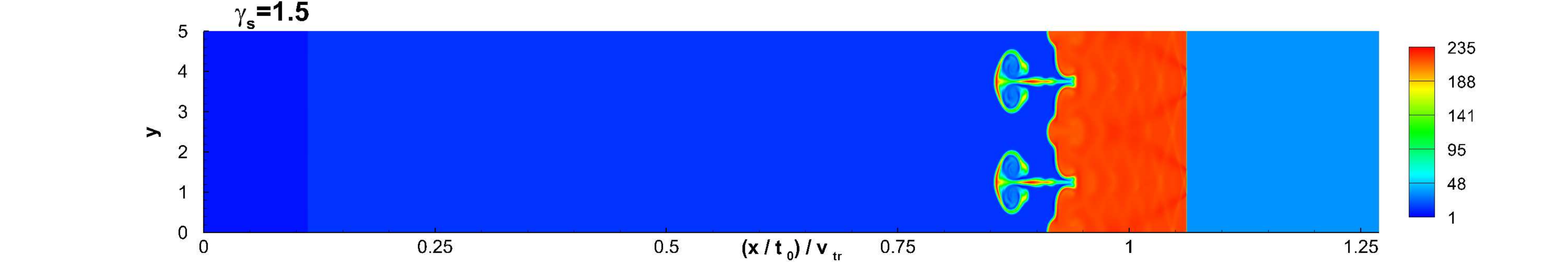}}
{\includegraphics[angle=0,width=12.0cm,height=1.9cm]{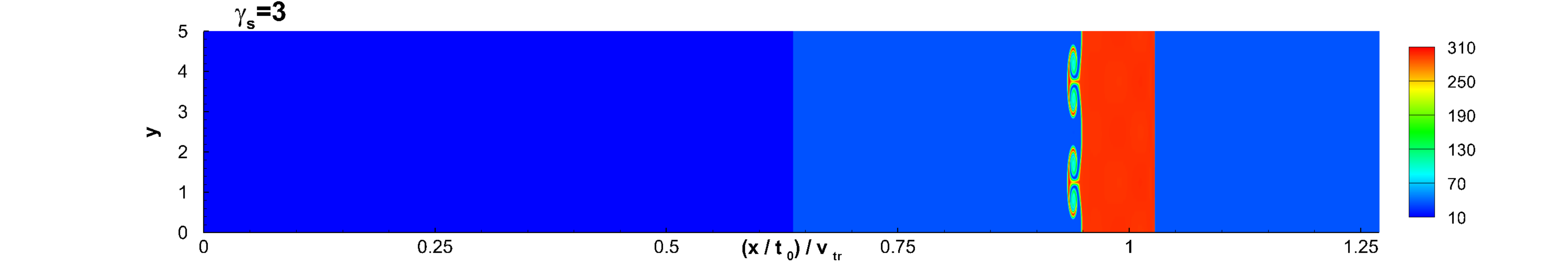}}
\end{center}
\caption{
{\em Light-to-Heavy models}.
Snapshot of the RM instability at time $t_0\sim 100$ for different values of the Lorentz factor $\gamma_{\rm s}$ and over a normalized horizontal grid. Note that $v_{tr}$ is the speed of the transmitted shock front.
}
\label{fig:RM-normalized}
\end{figure}

\begin{figure}
\begin{center}
{\includegraphics[angle=0,width=6.9cm,height=6.9cm]{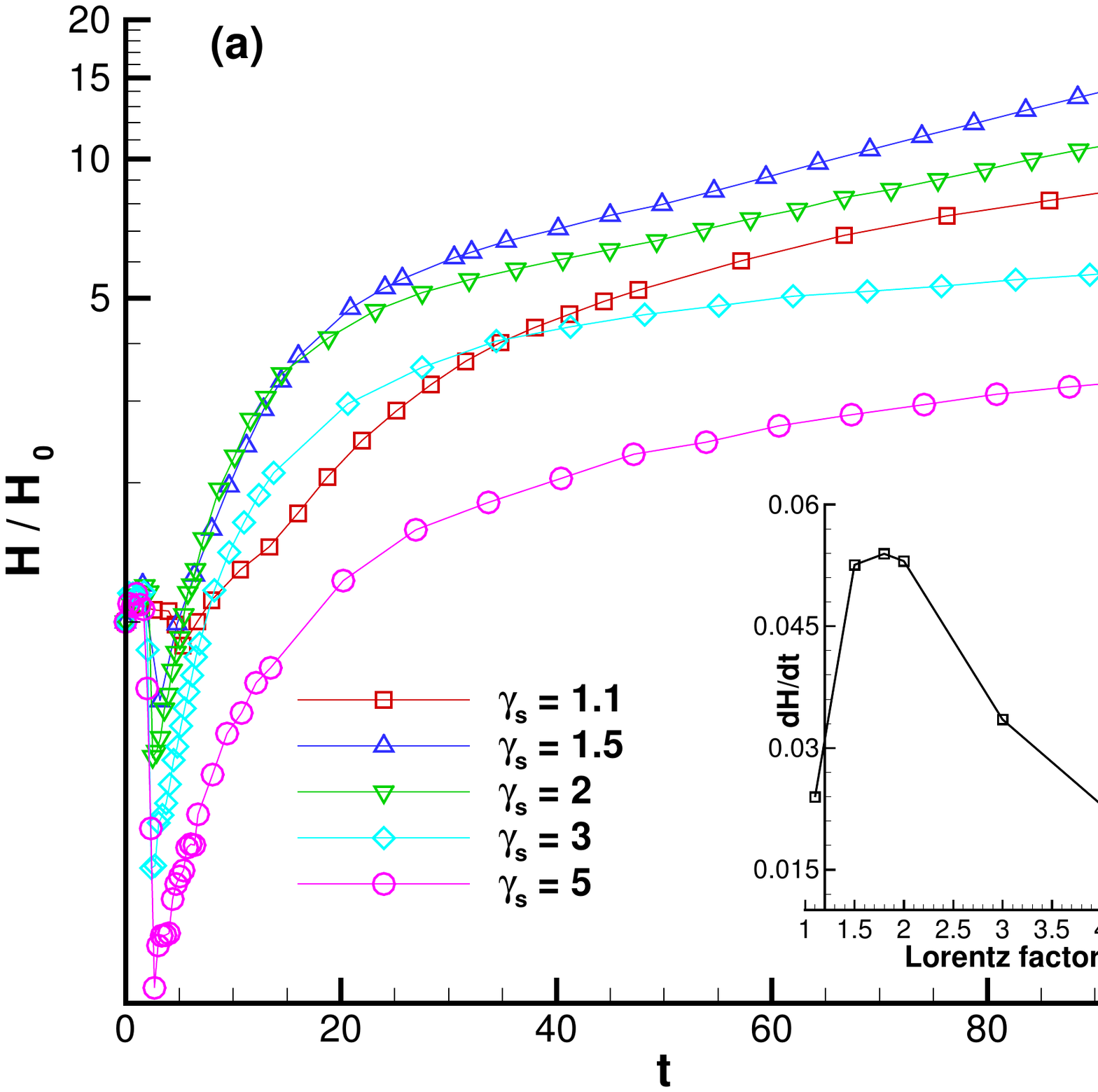}}
{\includegraphics[angle=0,width=6.9cm,height=6.9cm]{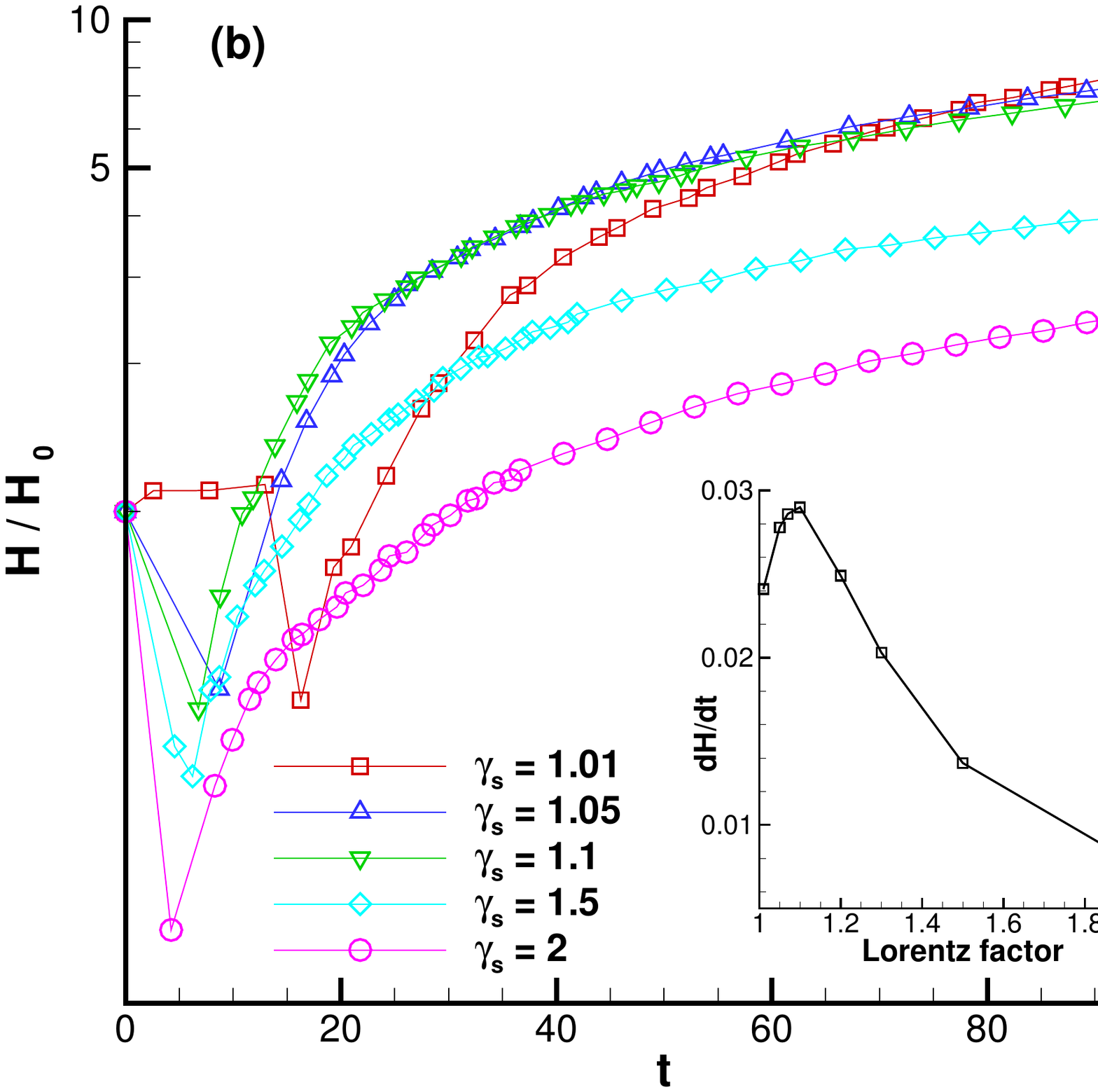}}
\end{center}
\caption{
Time evolution of the amplitude of the perturbation 
for different  Lorentz factors of the incident shock wave.
Panel (a): Light-to-Heavy models.
Panel (b): Heavy-to-Light models.
}
\label{fig:HversusTime}
\end{figure}

Soon after the incident shock wave hits the interface, the amplitude of the sinusoidal perturbation grows, generating
the typical "mushroom'' structures of the RM instability. 
Fig.~\ref{fig:RM-normalized} compares the development of the RM instability for three different values of the Lorentz factor $\gamma_{\rm s}$ at approximately the same time $t_0\approx 100$. The $x-$ axis have been normalized to $\tilde x = (x/t_0)/v_{tr}$, in such a way that the shocked region is around $\tilde x\sim1$ for every model.
Even from a  quick inspection, it is possible to conclude that, for large Lorentz factors $\gamma_{\rm s}$, the RM instability is largely reduced.
In order to make this statement quantitative, we have measured the amplitudes 
$H(t)$ of the growing sinusoidal perturbation
and we have monitored its time evolution for each model.\footnote{See \cite{Herrmann2007} for alternative diagnostics and tools to monitor the RM instability.} 
The results of our analysis are summarized in Fig.~\ref{fig:HversusTime}(a),
which shows the curves $H(t)/H_0$ for several values of $\gamma_{\rm s}$, where $H_0=2a=0.2$ is twice the initial amplitude of the sinusoidal perturbation. Each curve manifests
the same basic features:
there is an initial phase, just before the shock wave touches the sinusoidal interface, during which the amplitude $H(t)/H_0$ drops to very small values as a result of a local compression. After that, the sinusoidal interface is accelerated and the linear phase of the RM instability starts. The duration of the linear phase can vary from model to model, but it can be considered finished in all the models by the time $t\approx 20$.
A progressive transition to the non-linear regime then takes place, characterized by a substantially smaller growth rate, as reported by \citet{Holmes1999}.
%
%

Fig.~\ref{fig:HversusTime}(a) also highlights a subtle effect which may not be noticed at first sight. Namely, the amplitude $H(t)$ increases significantly 
in the  mild relativistic regime, i.e. when changing from $\gamma_{\rm s}=1.1$ to $\gamma_{\rm s}=1.5$. However, when $\gamma_{\rm s}$ is increased further, the RM instability is drastically reduced. 
This behaviour is confirmed by looking at the growth rate of the instability during the initial linear phase, which have been reported in the inset of Fig.~\ref{fig:HversusTime}(a) as a function of the Lorentz factor $\gamma_{\rm s}$. The inset clearly shows the existence of a critical Lorentz factor above which the RM instability becomes less and less efficient
and it is totally suppressed for $\gamma_{\rm s}=10$ (which has not been reported in the inset for visualization purposes). 
This result is in agreement with findings by \cite{Mohseni2013}. 
%

\begin{figure}[!htbp]
\begin{center}
{\includegraphics[angle=0,width=12.0cm,height=1.9cm]{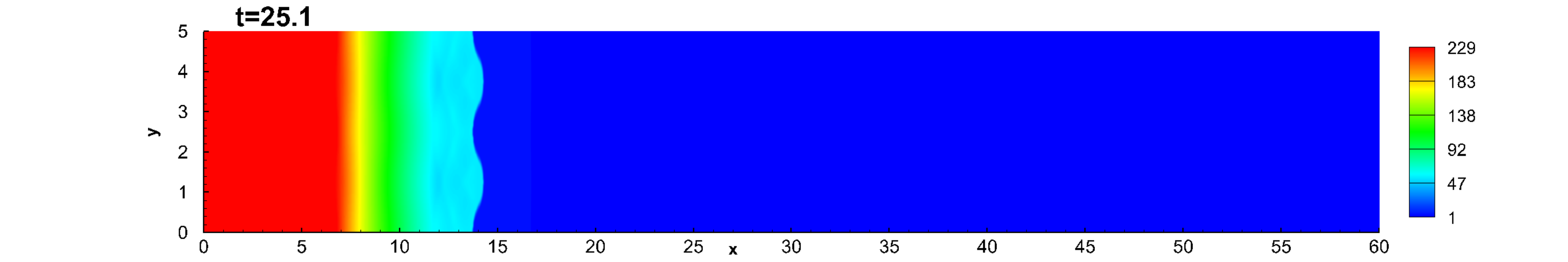}}
{\includegraphics[angle=0,width=12.0cm,height=1.9cm]{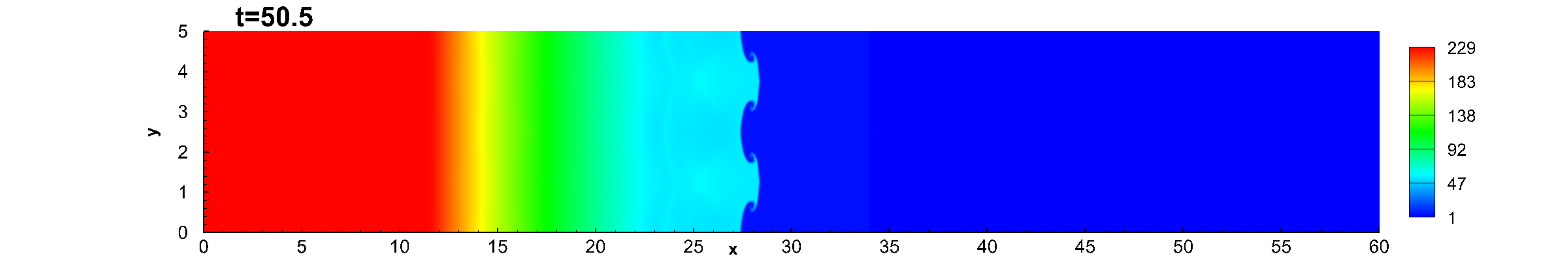}}
{\includegraphics[angle=0,width=12.0cm,height=1.9cm]{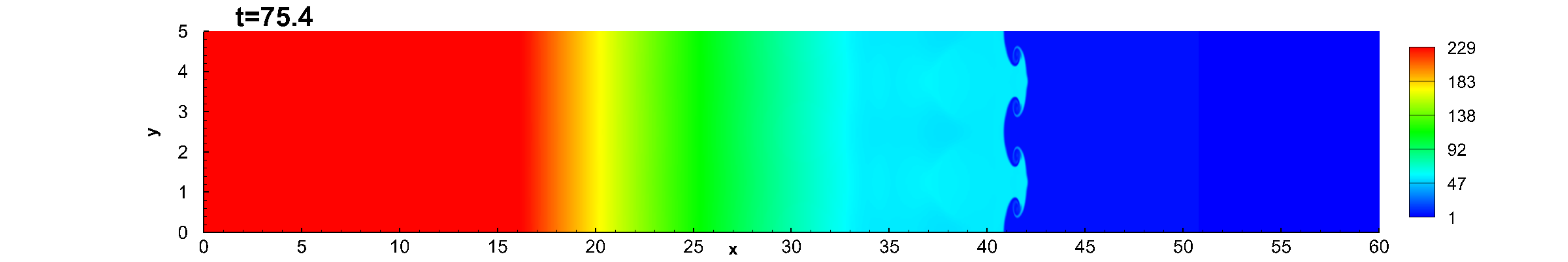}}
{\includegraphics[angle=0,width=12.0cm,height=1.9cm]{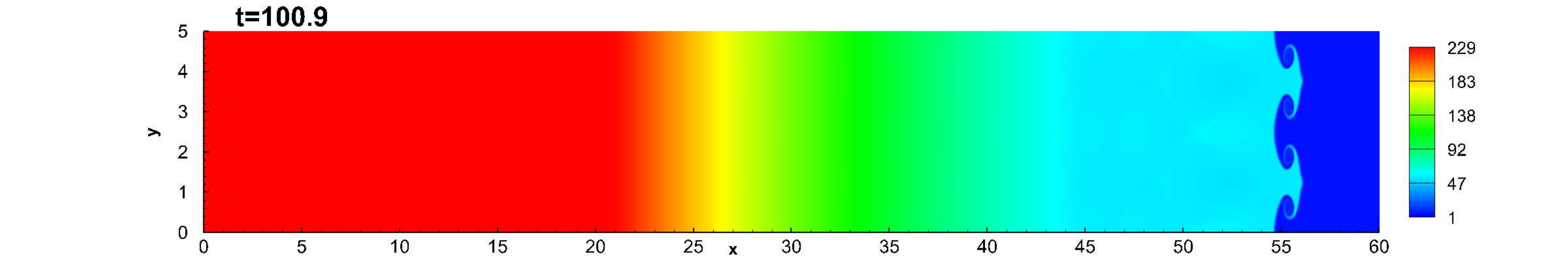}}
{\includegraphics[angle=0,width=12.0cm,height=1.9cm]{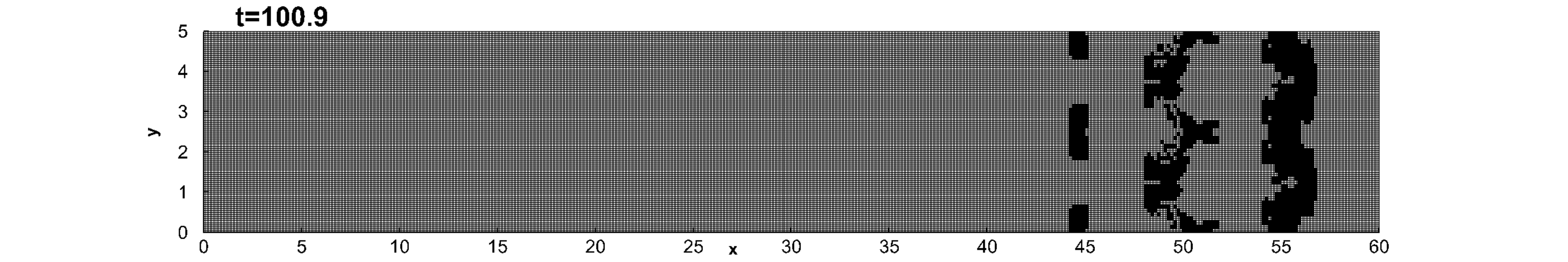}}
\end{center}
\caption{{\em Heavy-to-Light models}.
Time evolution of the RM instability in a configuration with Lorentz factor $\gamma_{\rm s}=1.1$ and $A=-0.94$.
At time $t=100.9$ the AMR grid is composed of $179,936$ elements.
}
\label{fig:RMW1.1-H2L}
\end{figure}
%
\begin{figure}
\begin{center}
{\includegraphics[angle=0,width=12.0cm,height=1.9cm]{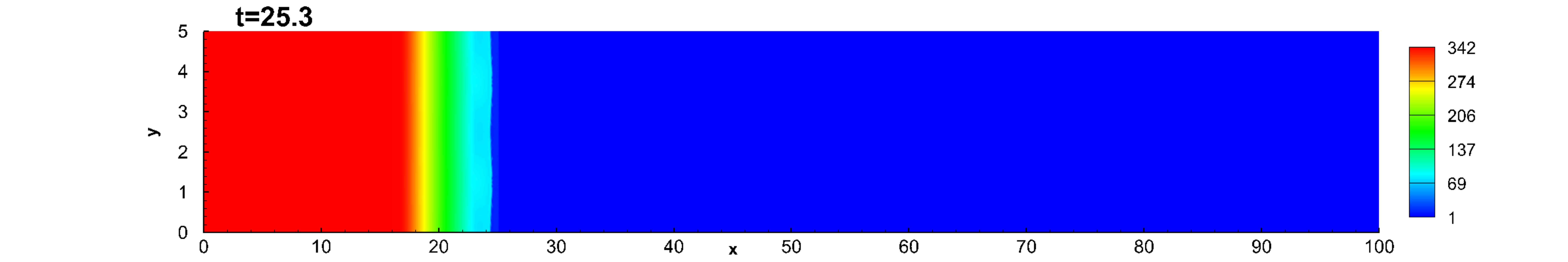}}
{\includegraphics[angle=0,width=12.0cm,height=1.9cm]{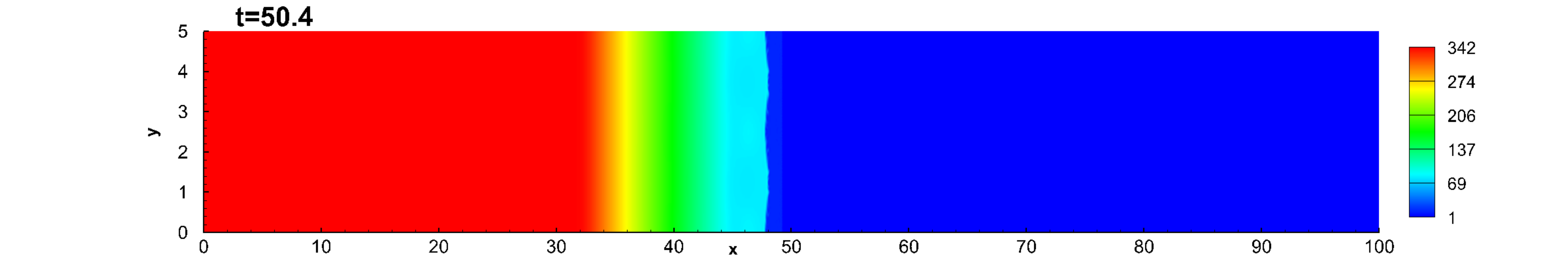}}
{\includegraphics[angle=0,width=12.0cm,height=1.9cm]{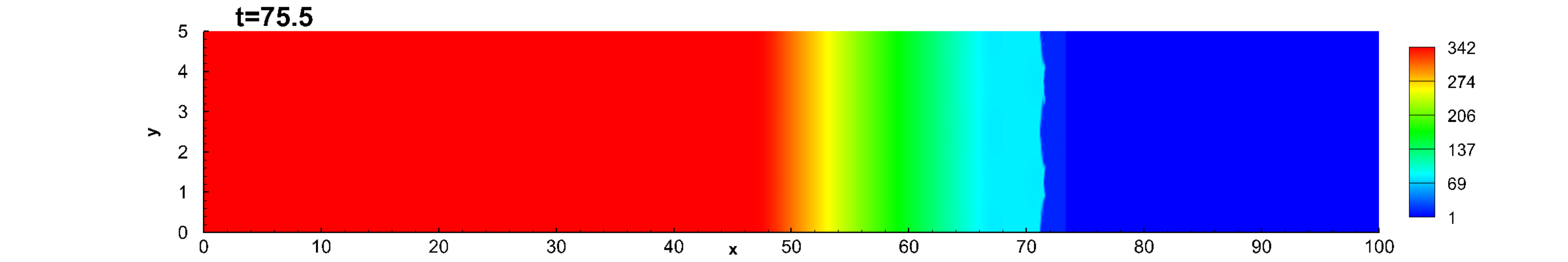}}
{\includegraphics[angle=0,width=12.0cm,height=1.9cm]{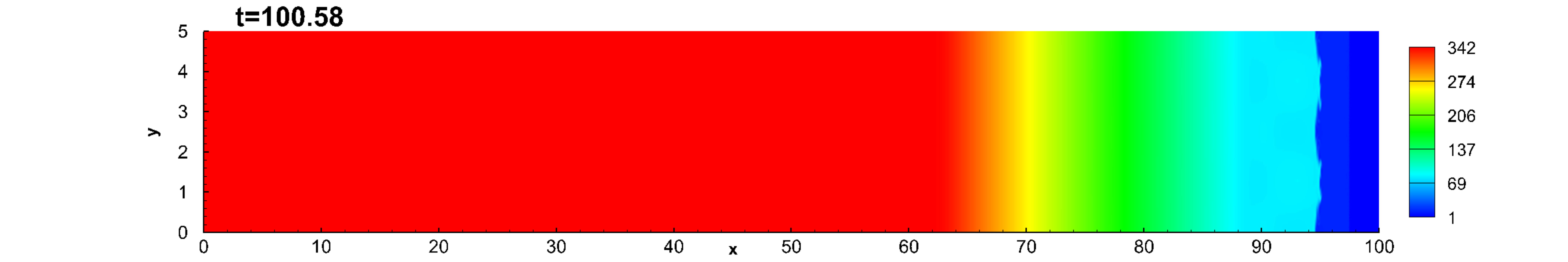}}
{\includegraphics[angle=0,width=12.0cm,height=1.9cm]{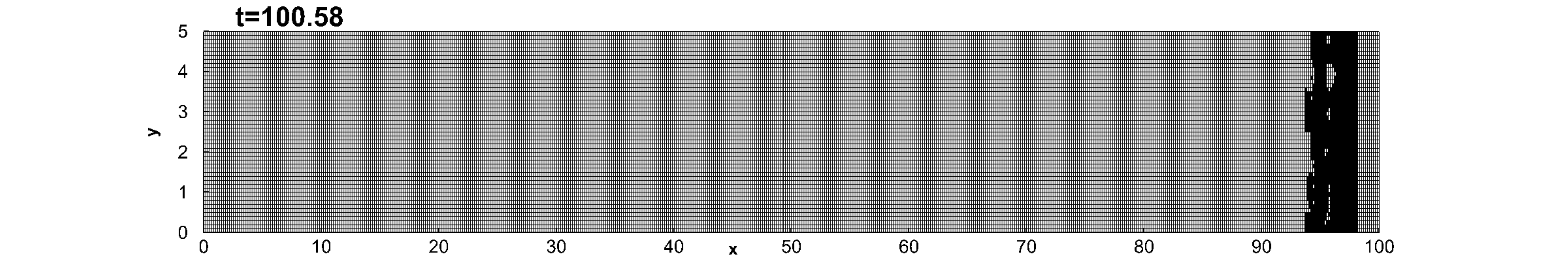}}
\end{center}
\caption{{\em Heavy-to-Light models}.
Time evolution of the RM instability in a configuration with Lorentz factor $\gamma_{\rm s}=2$ and $A=-0.94$.
At time $t=100.9$ the AMR grid is composed of $237,312$ elements.
}
\label{fig:RMW2-H2L}
\end{figure}
\begin{figure}
\begin{center}
{\includegraphics[angle=0,width=6.9cm,height=6.9cm]{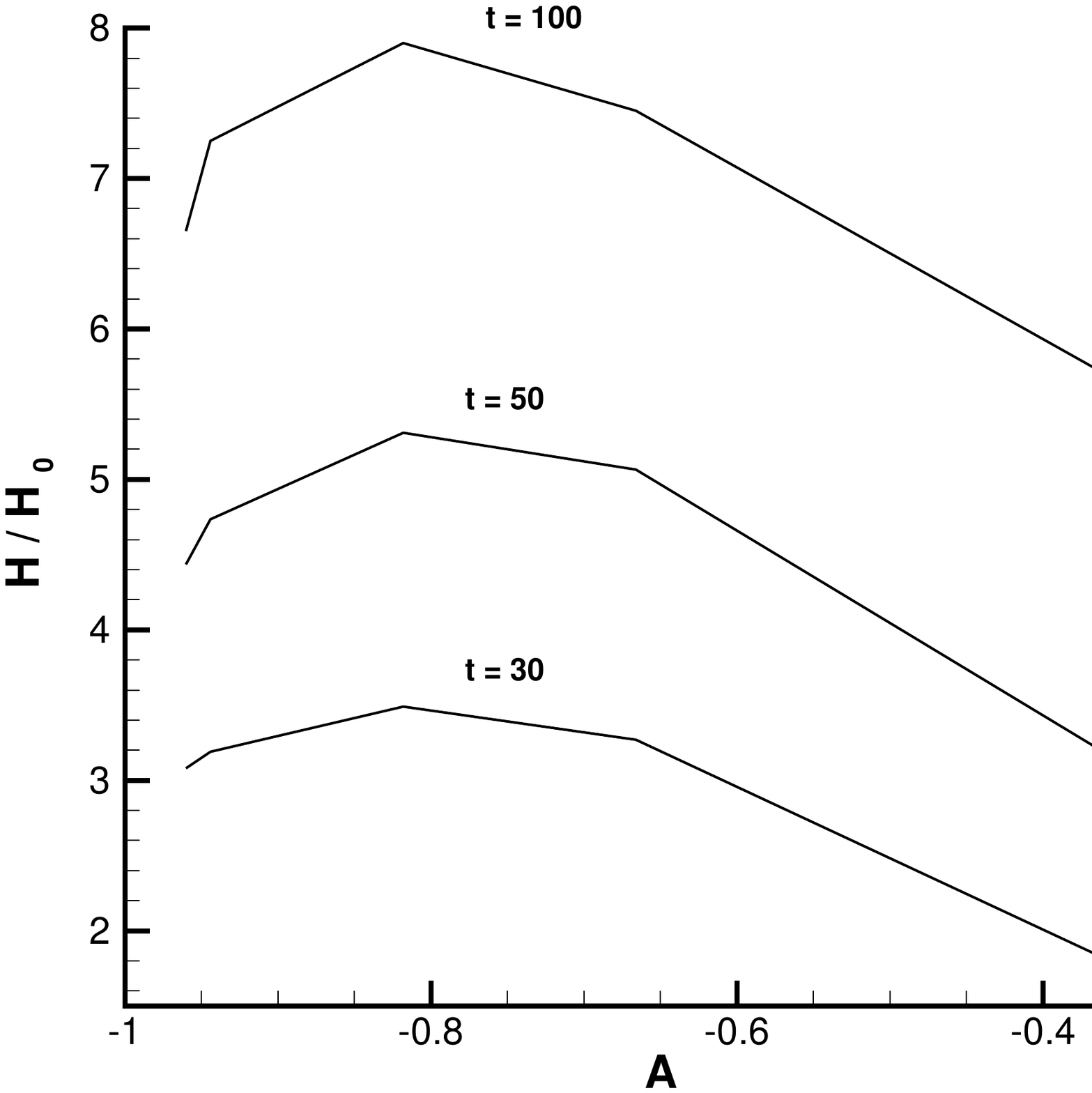}}
\end{center}
\caption{
Amplitude $H/H_0$, at time $t=30,50,100$, 
for different values of the Atwood number A in {\em heavy-to-light} models. 
The figure shows the existence of a local maximum at $A\sim -0.8$.
}
\label{fig:HversusAtwood}
\end{figure}
\subsubsection{Heavy-to-light models, $A<0$}
\label{sec:Heavy-to-light models}
In the second class of models that we have considered a high density fluid penetrates 
into a low density one, thus with a negative Atwood number.
We have repeated our analysis for this class, qualitatively confirming the basic results already described in Sect.~\ref{sec:Light-to-heavy models}
for the {\em Light-to-heavy} models. Indeed, Fig.~\ref{fig:RMW1.1-H2L} and Fig.~\ref{fig:RMW2-H2L} show that, by increasing the Lorentz factor of the incident shock wave
from $\gamma_{\rm s}=1.1$ to $\gamma_{\rm s}=2$, the RM instability is strongly weakened. The time evolution of the amplitude, $H(t)/H_0$, is also reported in 
Fig.~\ref{fig:HversusTime}(b), and is again characterized by the presence of an initial compression of the perturbation, followed by the linear growth of the RM instability, which becomes sub-linear around $t\sim 20$ for all the models considered. The inset of Fig.~\ref{fig:HversusTime}(b) allows to deduce the critical value of $\gamma_{\rm s}$ above which the linear growth rate of the RM instability decreases, which is actually quite small, i.e. $\gamma_{\rm s}\approx 1.15$.
By considering a few additional models, not reported in Table~\ref{tab:table1},
we have also monitored the dependence of the instability on the Atwood number, while keeping the Lorentz factor of the incident shock wave constant, i.e. $\gamma_{\rm s}= 1.1$. 
In particular, for $|A|\rightarrow 0$ and $|A|\rightarrow 1$ the instability is strongly weakened, while it has a maximum around $A\sim -0.8$. This effect is shown 
in Fig.~\ref{fig:HversusAtwood} which reports the profiles of $H/H_0$ as a function of $A$ at three different times.
%
%
\subsection{New relativistic effects on three dimensional configurations}
As anticipated in Sect.~\ref{sec:IC}, our main motivation for performing three-dimensional simulations of the relativistic RM instability is to verify whether any component of the velocity tangential to the shock front 
can affect the development of the instability. To this extent, we have first
compared the dynamics of the two {\em light-to-heavy} models
3D-Ia and 3D-Ib
(see Table~\ref{tab:table1}), for which only the latter has a $y-$component of the velocity. In fact, for the model 3D-Ib, $v_a^y=0.9$ all the way ahead of the shock wave, while $v_b^y$ is computed according to Eq.~(\ref{eq:Pons}). We emphasize that the non-zero tangential velocity is present in both the regions "L'' and 
"R'' of Fig.~\ref{fig:IC}, while the only jump of $v^y$ occurs across the impinging shock front. 
The pressure $p_b$ behind  the shock front is chosen in order to have $\gamma_{\rm s}=1.5$ 
as in model 3D-Ia.
The computational domain is $\Omega=[0,50]\times[0,5]\times[0,5]$, and it is initially covered by 
a level zero grid with $100 \times 28\times 28$ cells. Periodic boundary conditions are applied in the $y$ and $z$ direction.  
The refinement factor of the AMR algorithm is $\mathfrak{r}=2$, and two levels of refinement have been activated. This leads to 
an equivalent resolution on a uniform fine grid of $5,017,600$ cells.

\begin{figure}[!htbp]
\begin{center}
{\includegraphics[angle=0,width=6.5cm,height=6.5cm]{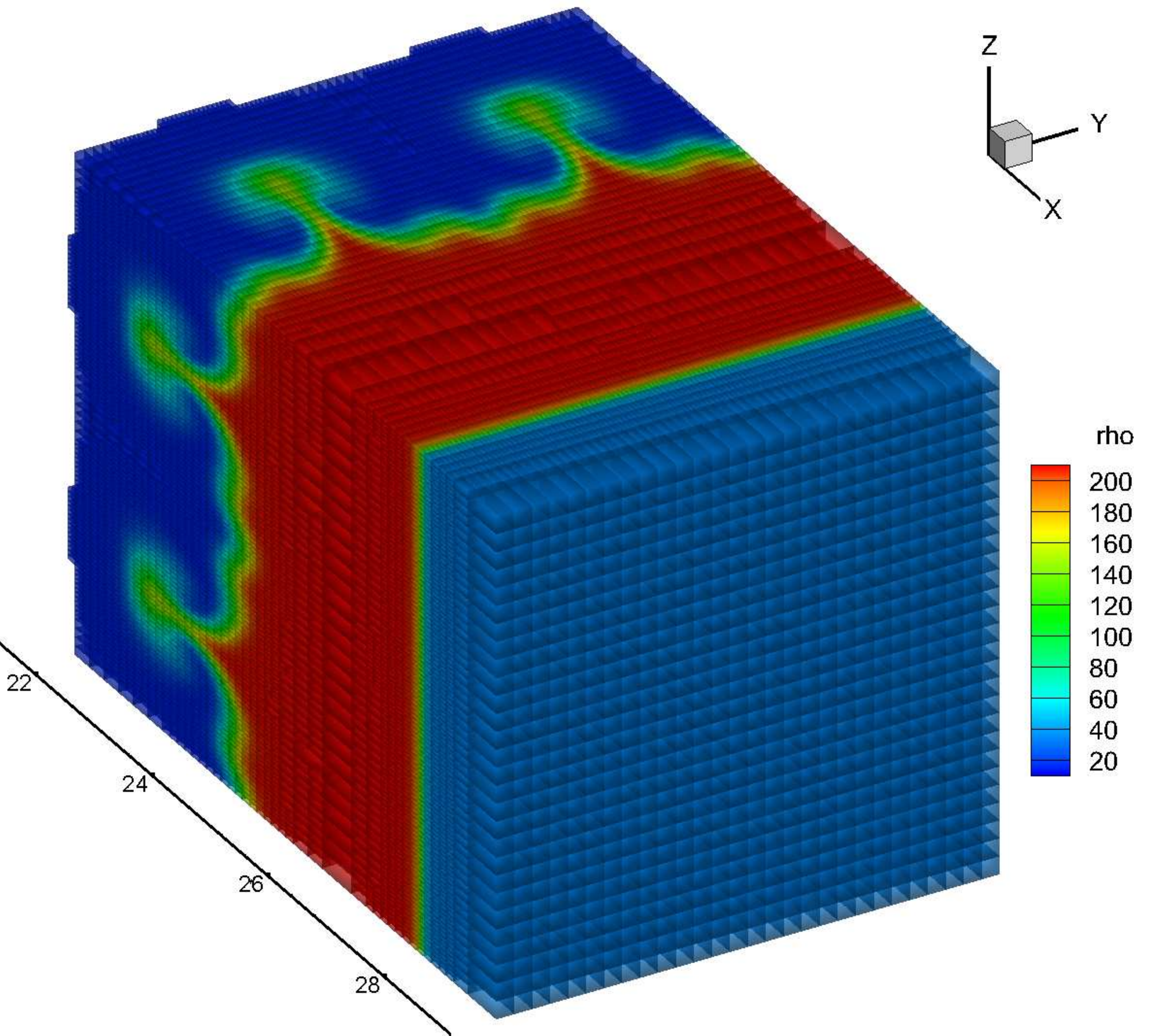}}
{\includegraphics[angle=0,width=6.5cm,height=6.5cm]{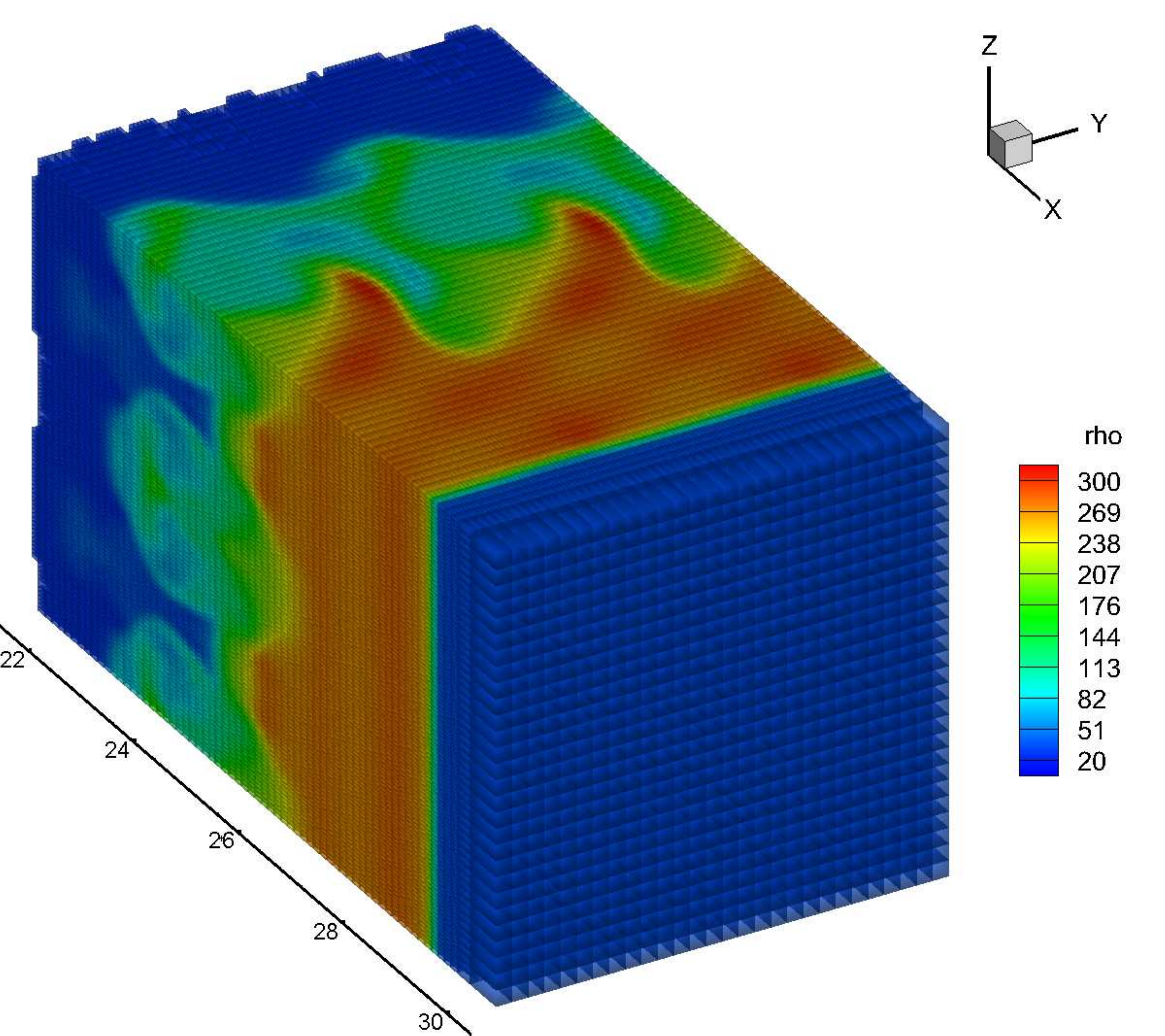}}
\end{center}
\caption{
RM instability in three dimensional calculations for the models 3D-Ia and 3D-Ib (light-to-heavy) at time $t=70$.
{\rm Left panel:}  Model 3D-Ia, without any velocity component tangential to the shock front.
{\rm Right panel:} Model 3D-Ib, with      a velocity component $v^t=0.9$ tangential to the shock front.
}
\label{fig:3D-Light-to-Heavy}
\end{figure}

\begin{figure}[!htbp]
\begin{center}
{\includegraphics[angle=0,width=6.5cm,height=6.5cm]{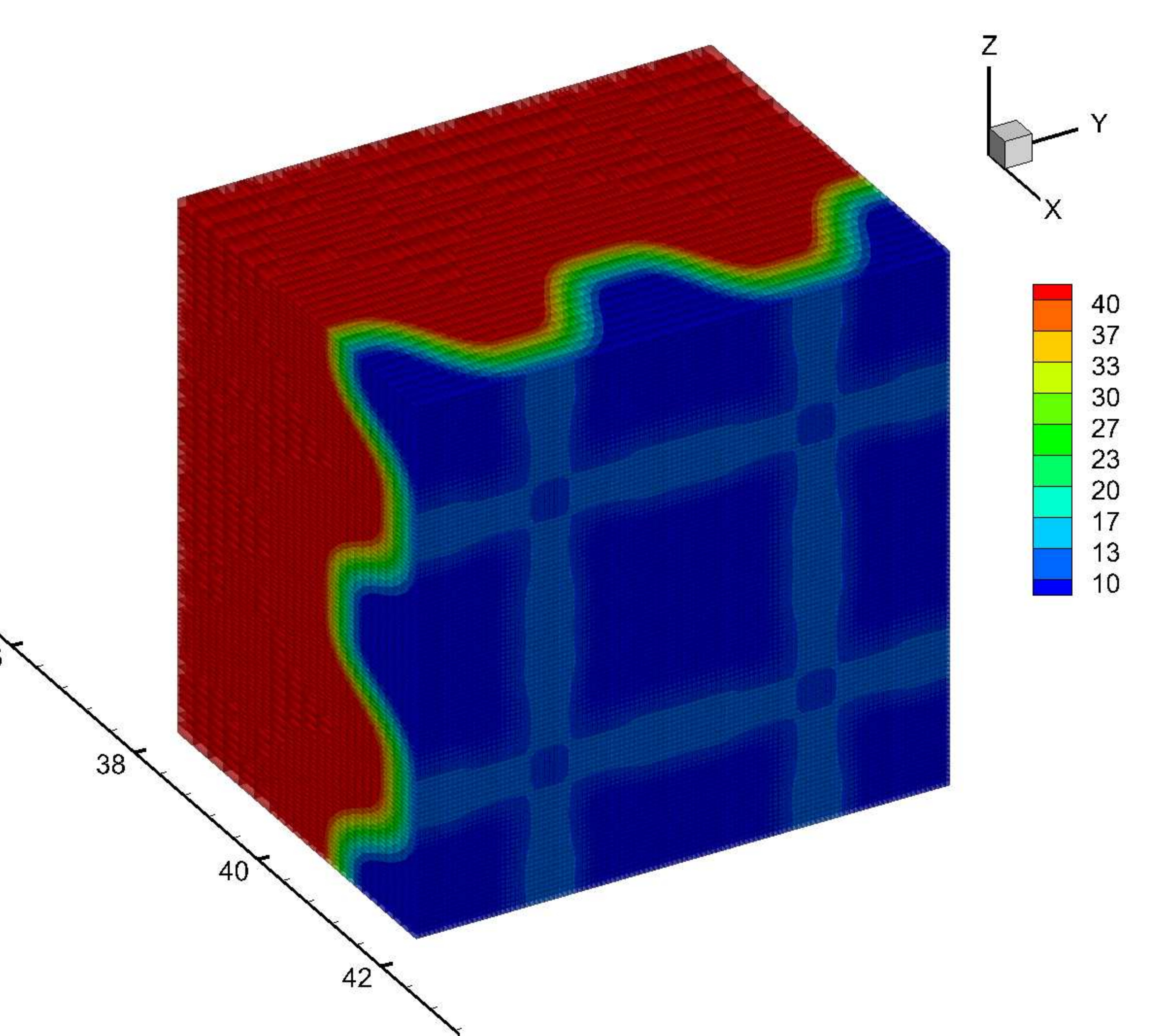}}
{\includegraphics[angle=0,width=6.5cm,height=6.5cm]{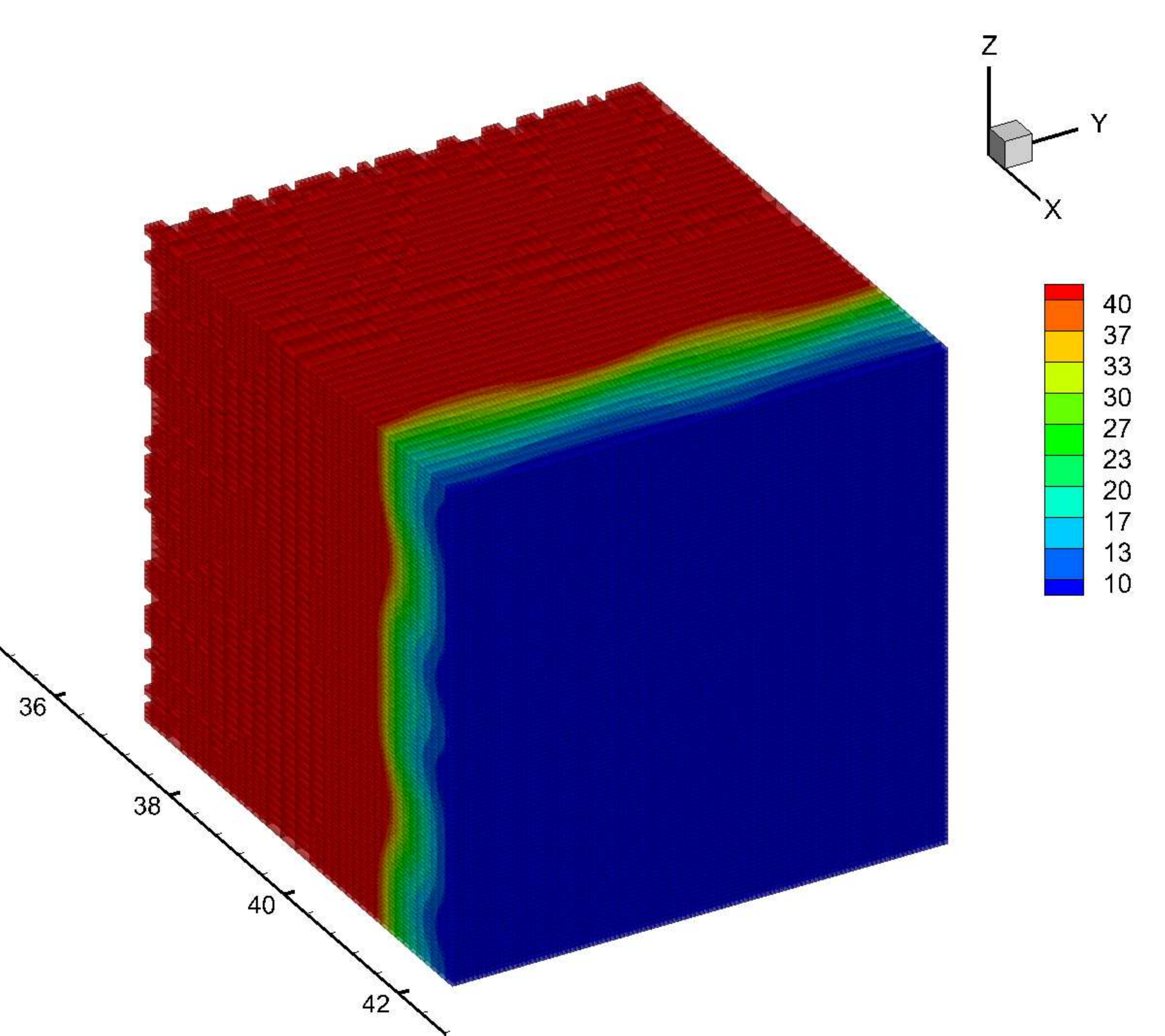}}
\end{center}
\caption{
RM instability in three dimensional calculations for the models 3D-IIa and 3D-IIb (heavy-to-light) at time $t=75$.
{\rm Left panel:}  Model 3D-IIa, without any velocity component tangential to the shock front.
{\rm Right panel:} Model 3D-IIb, with      a velocity component $v^t=0.9$ tangential to the shock front.
}
\label{fig:3D-Heavy-to-Light}
\end{figure}

We recall that in Newtonian hydrodynamics a tangential velocity $v^t$ produces a simple shift of the "mushroom'' structures of the RM instability, but 
it does not affect the development of the instability in any other way. 
The results are totally different in the relativistic regime and they are reported in Fig.~\ref{fig:3D-Light-to-Heavy}, which shows a zoom into the numerical domain of interest.
The left panel corresponds to $v^t=0$, hence with no tangential velocity, while the right panel refers to the model with $v^t=v^y=0.9$. The time is $t=70$ in the two cases.
 As it is apparent from the figure, the effect of the tangential velocity is to significantly modify the development  of the RM instability. In particular, the size of the shocked region is reduced, and the penetration of the high density fluid into the lower density one is less sharp. 
We have repeated a similar comparison also for two  representative {\em heavy-to-light} models, 3D-IIa and 3D-IIb
(see Table~\ref{tab:table1}), for which only the latter has a $y-$component of the velocity. The results are shown in Fig.~\ref{fig:3D-Heavy-to-Light} and in this case the effect of the tangential velocity is even more drastic, since the RM instability is essentially suppressed (see right panel).

Although there is not an intuitive explanation of these results, which are highly non-linear and can only be revealed by means of numerical simulations,
they are not totally surprising. In relativistic hydrodynamics, in fact, the tangential velocities (either along the $y$ or the $z$ direction) are not continuous across the shock front and, by coupling non-linearly through the Lorentz factor with the normal component of the velocity (along the $x$ direction), they can strongly modify the dynamics of the flow, especially if a fluid instability is present. 
In a simpler physical set-up, namely in the solution of the relativistic Riemann problem,
analogous dynamical effects due to non-zero tangential velocities 
were already reported by  \cite{Rezzolla02}.

%

\section{Conclusions}
\label{sec:conclusions}
We have performed numerical simulations of the relativistic Richtmyer--Meshkov (RM) instability both in two and in three dimensional configurations.
The numerical scheme used in this study adopts a high order accurate one-step finite volume discretization. In addition, space-time adaptive mesh 
refinement (AMR) has been used, which is crucial to resolve the finest flow details of the RM instability. To our knowledge, this is the first 
time that a better than second order accurate finite volume scheme with space-time adaptive mesh refinement has been applied to this kind of 
physical problem.  
By investigating a wide parameter space, we have shown that by increasing the Lorentz factor $\gamma_{\rm s}$ of the incident shock wave, the RM 
instability is progressively attenuated, both in models with Atwood number $A>0$, ({\em light-to-heavy} models), as well as in models with 
$A<0$, ({\em heavy-to-light} models). 
In the two classes of models the growth rate of the RM instability in the linear phase has a local maximum that occurs at 
a critical value of $\gamma_{\rm s}\approx [1.2,2]$. 
Finally, when a velocity component tangential to the shock front is introduced, the RM instability is strongly affected, marking a strong difference 
with respect to classical Newtonian hydrodynamics. In particular, we have found that  a tangential velocity $v^t=0.9$ produces a 
significant distortion of the typical RM phenomenology and a less efficient mixing of the fluid. This result may have important implications in 
relativistic astrophysical jets, for which the RM instability is supposed to develop at the interface between the jet and the surrounding medium. 
In such circumstances, the bulk motion along the axis of the jet acts as a tangential velocity, possibly reducing the role of the RM instability 
in determining the transverse structure.

\section*{Acknowledgments}
The research presented in this paper has been financed by the European Research Council (ERC) under the
European Union's Seventh Framework Programme (FP7/2007-2013) with the research project \textit{STiMulUs}, 
ERC Grant agreement no. 278267.\\
The authors would like to acknowledge PRACE for awarding access to the SuperMUC 
supercomputer based in Munich, Germany at the Leibniz Rechenzentrum (LRZ), and ISCRA, for awarding access to the FERMI supercomputer based in Casalecchio (Italy).

\bibliographystyle{aipauth4-1}
\bibliography{aeireferences}


\end{document}